\documentclass[]{article}
\usepackage[a4paper, total={7in, 9in}]{geometry}
\usepackage{amsmath}
\usepackage{graphicx}
\usepackage{float}
\usepackage{xcolor}
\usepackage{graphicx}
\usepackage{dcolumn}
\usepackage{bm}
\usepackage{subcaption}
\usepackage{caption}
\usepackage{multicol}
\usepackage{authblk}
\usepackage[T1]{fontenc}

\usepackage{hyperref}
\hypersetup{colorlinks=true, citecolor=blue, urlcolor=blue, linkcolor=blue}

\def \etal {{\it et al.}}

\title{Reduction of coherent betatron oscillations in a muon g-2 storage ring experiment using RF fields}
\author[1,2]{On Kim}
\affil[1]{Institute for Basic Science, Daejeon 34051,  South Korea}
\affil[2]{Korea Advanced Institute for Science and Technology, Daejeon 34141, South Korea}
\author[3]{Meghna Bhattacharya}
\affil[3]{University of Mississippi, University, MS 38677, USA}
\author[1,2]{SeungPyo Chang}
\author[1]{Jihoon Choi}
\author[4,5]{Jason D. Crnkovic}
\affil[4]{University of Illinois, Urbana, IL  61820, USA}
\affil[5]{Brookhaven National Laboratory, Upton, NY 11973, USA}
\author[4]{Sudeshna Ganguly}
\author[1]{Selcuk Hac\i\" omero\u glu\thanks{Corresponding author:selcuk.haciomeroglu@gmail.com}}
\author[6]{Manolis Kargiantoulakis}
\affil[6]{Fermi National Accelerator Laboratory, Batavia, IL 60510, USA}
\author[1,7]{Young-Im Kim }
\affil[7]{Korea University,  Seoul 136-701, South Korea}
\author[1]{Soohyung Lee}
\author[5]{William M. Morse}
\author[6]{Hogan Nguyen}
\author[8]{Yuri F. Orlov}
\affil[8]{Cornell University, Ithaca, NY 14853, USA} 
\author[9]{B. Lee Roberts}
\affil[9]{Boston University, Boston, MA 02215 USA}
\author[1,2]{Yannis K. Semertzidis}
\author[5]{Vladimir Tishchenko}
\author[9]{Nam H. Tran}
\author[4]{Esra Barlas Yucel}

\begin{document}
\onecolumn
\maketitle


\begin{abstract}
This work demonstrates that two systematic errors, coherent betatron oscillations (CBO) and muon losses can be reduced through application of radio frequency (RF) electric fields, which ultimately increases the sensitivity of the muon $g-2$ experiments. As the ensemble of polarized muons goes around a weak focusing storage ring, their spin precesses, and when they decay through the weak interaction, $\mu^+ \rightarrow e^+ \nu_e  \bar{\nu_\mu}$, the decay positrons are detected by electromagnetic calorimeters. In addition to the expected exponential decay in the positron time spectrum,  the weak decay asymmetry causes a modulation in the number of positrons in a selected energy range at the difference frequency between the spin and cyclotron frequencies, $\omega_\text{a}$. This frequency is directly proportional to the magnetic anomaly $a_\mu =(g-2)/2$, where $g$ is the g-factor of the muon, which is slightly greater than 2.  The detector acceptance depends on the radial position of the muon decay, so the CBO of the muon bunch following injection into the storage ring modulate the measured muon signal with the frequency $\omega_\text{CBO}$. In addition, the muon populations at the edge of the beam hit the walls of the vacuum chamber before decaying, which also affects the signal. Thus, reduction of CBO and unwanted muon loss increases the $a_\mu$ measurement sensitivity. Numerical and experimental studies with RF electric fields yield more than a magnitude reduction of the CBO, with muon losses comparable to the conventional method.
\end{abstract}
\twocolumn
\section{Introduction}
Efforts to measure the muon magnetic anomaly $a_\mu\equiv(g_\mu-2)/2$ using storage rings have been ongoing since the 1960s \cite{ref:miller_2007}. Several key methods like the magic momentum were applied in experiments at CERN \cite{ref:bailey_1969pr,ref:cern_3, ref:cern_4}. The use of electrostatic quadrupoles was first introduced in the third CERN experiment,  which obtained a precision of 7.3~ppm~\cite{ref:cern_4}.  With a significant increase in the number of muons and direct muon injection into the storage ring, the Brookhaven  National Laboratory (BNL) experiment E821 achieved 0.54 ppm precision~\cite{ref:bnl_final_report}. At present, there appears to be a greater than 3 standard deviation difference between the experimental measurement of $a_\mu$ and the Standard Model value \cite{ref_sensitivity_1, ref:sensitivity_2}. Currently, the Fermilab $g-2$ collaboration is conducting a new experiment (E989) with upgraded muon
beamline,  storage ring and detector systems~\cite{ref:fermilab}. The E989 goal is to measure $a_\mu$ to an accuracy of 140 ppb, four times the precision reached by BNL E821.

The positive muon beam in E989 is produced in two steps: First, a proton beam hits a target to produce copious pions that decay into muons and neutrinos. After the pions decay, a small muon momentum bite is selected, yielding a longitudinally polarized ($>95\%$) muon beam at the magic momentum ($p_\text{m}=3.09 ~\text{GeV}/c$) that is then injected into the 14 m diameter storage ring. The beam fills the  ring azimuthally in the first turn, being stored radially by a $B_0=1.45$ T vertical dipole magnetic field and vertically by four sets of electrostatic quadrupoles.  

The storage ring magnet is designed as a near continuous ``C'' magnet with yoke sections that have minimal spaces between them~\cite{ref:Danby_2001eh}. This design is required in order to shim the dipole component to high uniformity in
azimuth. After the magnetic shimming at Fermilab, the dipole component has achieved an RMS of $\pm 15$~ppm over its 44.7 m circumference inside of the 9~cm diameter muon storage region.  The fast muon kicker and the electrostatic quadrupoles residing inside the beam vacuum chamber are constructed from non-ferromagnetic materials. 

\begin{figure}
	\centering
	\includegraphics[width=\linewidth]{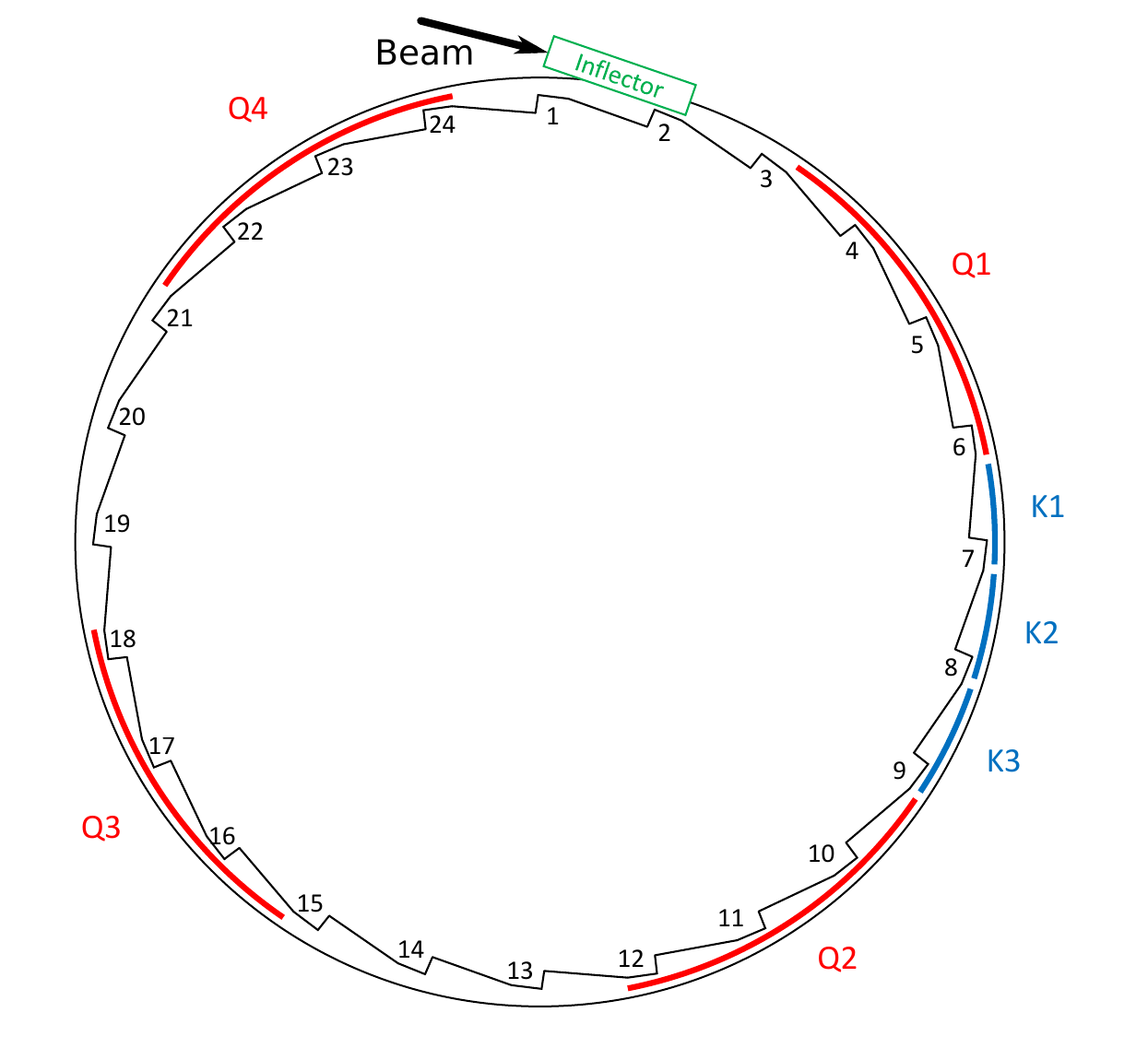}
	\caption{A plan view of the E989 7.1 m radius storage ring vacuum chamber. ``Q'' and ``K'' indicate quad and kicker sections, respectively. While not shown in the figure, each quadrupole has long (3.2 m) and short (1.6 m) sections. The beam is injected through a field-free inflector section located 7.7 cm outside the main magnetic field. Then, it crosses the design orbit at around the beginning of Section 7. The three kicker sections provide a radial kick to the beam to put it onto the design orbit.}
	\label{fig:ring_from_top}
\end{figure}

\begin{figure}
	\centering
	\includegraphics[width=0.8\linewidth]{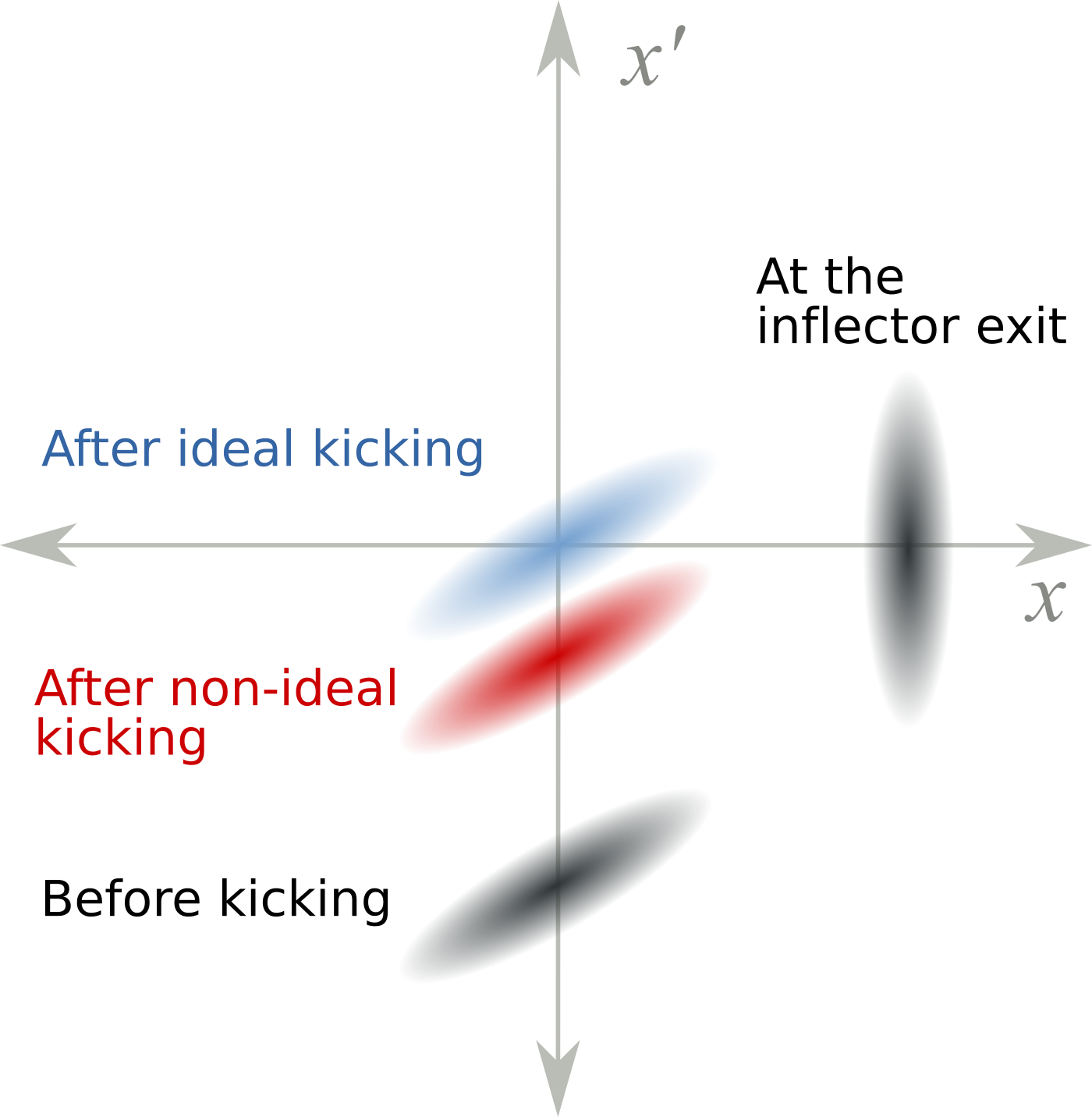}
	\caption{Illustration of the radial phase space of the incoming beam.  The incoming beam rotates in phase space and then is kicked into the acceptance of the storage ring by the kicker magnets. The blue (ideal) and red (achieved) distributions after the kick are shown. The beam is underkicked, resulting in a large coherent betatron oscillation.}
	\label{fig:kicker_at_phase_space}
\end{figure}

Figure \ref{fig:ring_from_top} shows the top view of the storage ring. The beam is brought into the storage region through a superconducting septum magnet called the inflector~\cite{ref:Yamamoto_2002bb}. After the beam exits the inflector magnet, it is kicked onto a stable orbit with a fast kicker~\cite{Schreckenberger:2018njd}. The kicker provides a 200 Gauss vertical field with a 200 ns width. It lasts around 1 $\mu$s with ringing. The limited space between the storage ring pole pieces necessitates a very narrow beam channel through the inflector magnet, making it impossible to match the incident beam phase space with that of the storage ring. The present strength of the kicker magnet is slightly less than needed to center the beam in the storage region. As one can infer from Figure \ref{fig:kicker_at_phase_space}, the mismatch of the phase space combined with the underkick leads to a significant radial beam oscillation after injection. It has a decoherence time of 200-300 $\mu$s, which is not insignificant compared to the dilated muon lifetime of 64.4~$\mu$s. This radial oscillation is called the ``coherent betatron oscillation'' (CBO) described below.

Assuming that $\vec B$ is uniform and $\vec \beta \perp \vec B$, the difference between this muon spin frequency and the cyclotron frequency is:
\begin{eqnarray}
\vec \omega_{a_\mu} &=& \vec \omega_\text{S} - \vec \omega_\text{C} \\
&=& -g_\mu\frac{q \vec B}{2m_\mu} -
(1-\gamma)\frac{q \vec B}{\gamma m_\mu}+ \frac{q \vec B}{\gamma m_\mu}\\
&=&
- \left(\frac{g_\mu-2}{2}\right) \frac{q}{m_\mu} \vec B
= - a_\mu\frac{q}{m_\mu} \vec B \, ,
\label{eq:diffreq}
\end{eqnarray}
where the muon charge $q = \pm \vert e \vert$. This difference frequency  provides a direct measurement of the magnetic anomaly.  With the presence of the electric quadrupole field, and assuming that $\vec \beta \perp \vec B$ and $\vec \beta \perp \vec E$, the spin equation is modified to~\cite{ref:cern_4}
\begin{equation}
\vec \omega_\text{a} \approx  - \frac{q}{ m} \left[ a_{\mu} \vec B -
\left( a_{\mu}- {1 \over \gamma^2 - 1} \right) 
{ {\vec \beta \times \vec E }\over c }\right]\,.
\end{equation}
For $\gamma_{magic} = 29.3$ ($p_{magic}=3.09$~GeV/c) the electric field does not contribute to the rate at which the spin rotates relative to the momentum~\cite{Miller:2018jum}. After approximately 30 turns around the ring, the spin angle relative to the momentum vector turns through $2\pi$.

When the stored muons decay, the highest energy positrons in the muon rest frame are correlated with the muon spin. As the muon spin precesses relative to the momentum vector, the number of high-energy decay positrons observed in the lab frame oscillates with the frequency $\omega_\text{a}$. This feature is exploited by selecting the highest energy positrons in the calorimeters and measuring the muon spin precession rate in the horizontal plane.

\begin{figure}
	\centering
	\includegraphics[width=\linewidth]{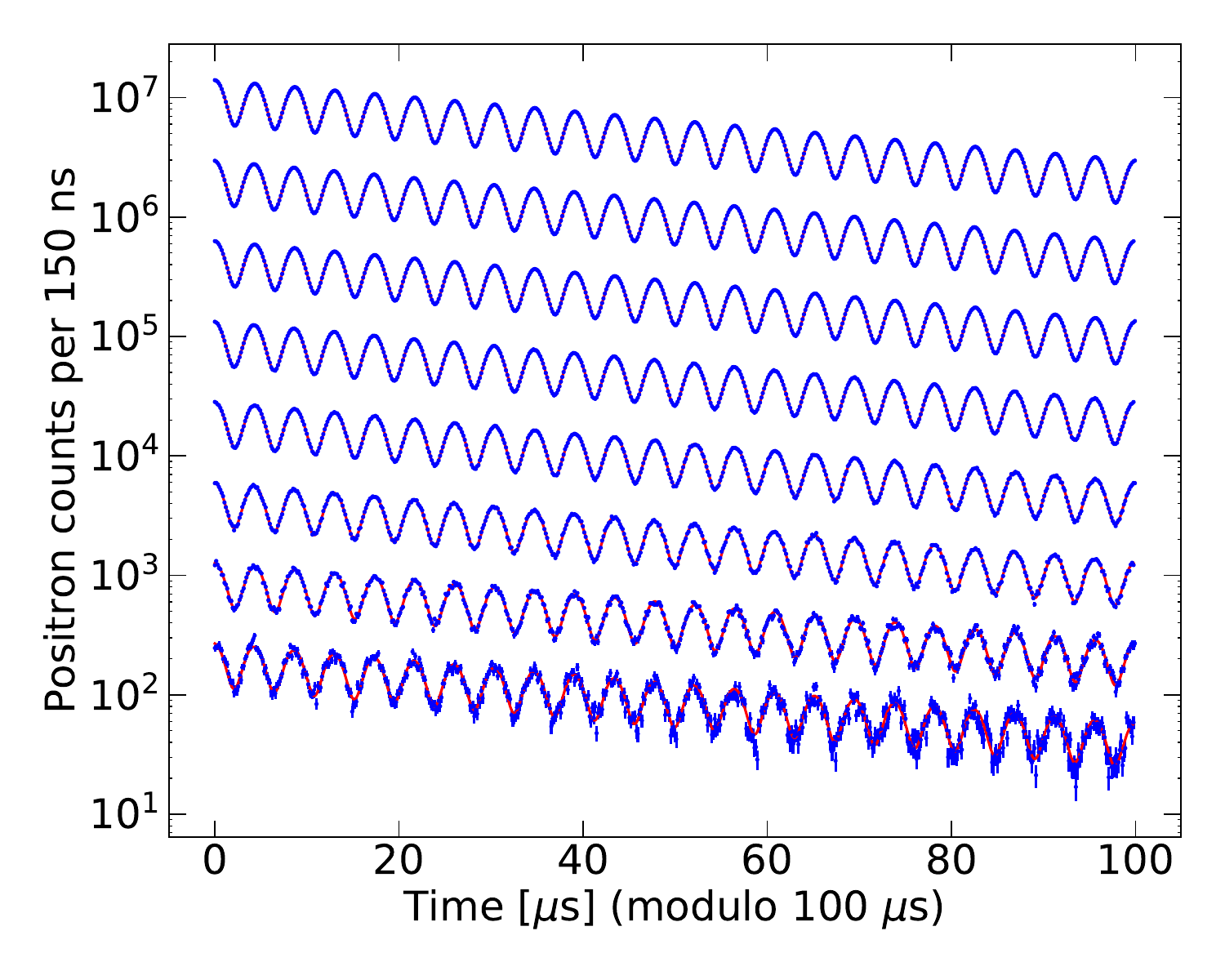}
	\caption{The arrival time spectrum of high-energy electrons as estimated with a Monte Carlo simulation. The blue points are data and the red curve is the fit to the data. The histogram
		contains $4.3\times10^9$ events.} 
	\label{fig:wiggle01}
\end{figure}

A sample time spectrum of high-energy electrons is shown in Figure \ref{fig:wiggle01}. These data appear to show the muon lifetime modulated by the muon spin frequency $\omega_\text{a}$, which can be represented by the "five-parameter function" 
\begin{equation}
N(t)=N_0 e^{-t/{\gamma \tau_\mu}}[1+A\cos(\omega_\text{a} t + \phi)] ,
\label{eq:five_param}
\end{equation}
where $N_0$ is the number of particles per 149~ns time bin at $t=0$, $\gamma=29.3$ is the relativistic Lorentz factor, $\tau_\mu\approx2.2~\mu$s is the lifetime of muon at rest, $A$ is the overall decay asymmetry, and $\phi$ is the $g-2$ phase. However, there is also information in these data on the coherent beam motion discussed below. 

With a continuous pure electric quadrupole field around the ring to provide vertical focusing, the transverse motion of the beam follows the linear Hill's equations:
\begin{eqnarray}
\label{eq:eq_of_mot_x}
& \frac{\mathrm{d}^2 x}{\mathrm{d} t^2} &+ \omega_\text{c}^2 (1-n) x = 0, \\
& \frac{\mathrm{d}^2 y}{\mathrm{d} t^2} &+ \omega_\text{c}^2 n y = 0,
\end{eqnarray}
where $\omega_\text{c}/2\pi \simeq 6.7$ MHz  is the cyclotron frequency, and $n \simeq 0.12$ is the weak focusing field index, determined by the ring radius $R_0$, the magnetic field $B_0$, the particle velocity $v$, and the electric quadrupole gradient $\kappa$: 
\begin{equation}
n=\frac{\kappa R_0}{v B_0}.
\end{equation}
The betatron frequencies in the horizontal and vertical directions are $\omega_\text{x}=\omega_\text{c}\sqrt{1-n}$ and $\omega_\text{y}=\omega_\text{c}\sqrt{n}$, respectively. 

\begin{figure}[h!]
		\centering
	\includegraphics[width=0.8\linewidth]{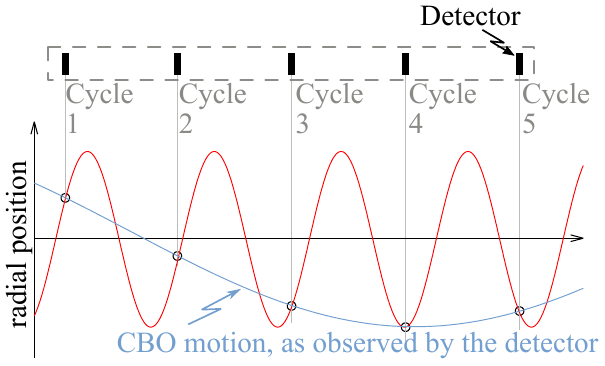}
	\caption{Illustration of the radial beam motion in the storage ring, showing successive turns in the ring. The betatron oscillation (red) frequency is adjusted to be less than the cyclotron frequency. A detector sees the motion at the CBO frequency. The location of a single detector is shown.}
	\label{fig:cbobetalambda}
\end{figure}

The horizontal betatron oscillation frequency is slightly less than the cyclotron frequency. Therefore, a localized detector sees
the beam oscillate inward and outward at $f_\text{CBO}\simeq0.4$ MHz, namely the CBO frequency~\cite{ref:bnl_final_report}, defined as  
\begin{equation*}
2\pi f_\text{CBO}=\omega_\text{CBO}=\omega_\text{c} - \omega_\text{x}=\omega_\text{c}(1-\sqrt{1-n}).
\end{equation*}
An illustration of this motion is shown in Figure \ref{fig:cbobetalambda}.

\begin{figure}[h!]
		\centering
	\includegraphics[width=\linewidth]{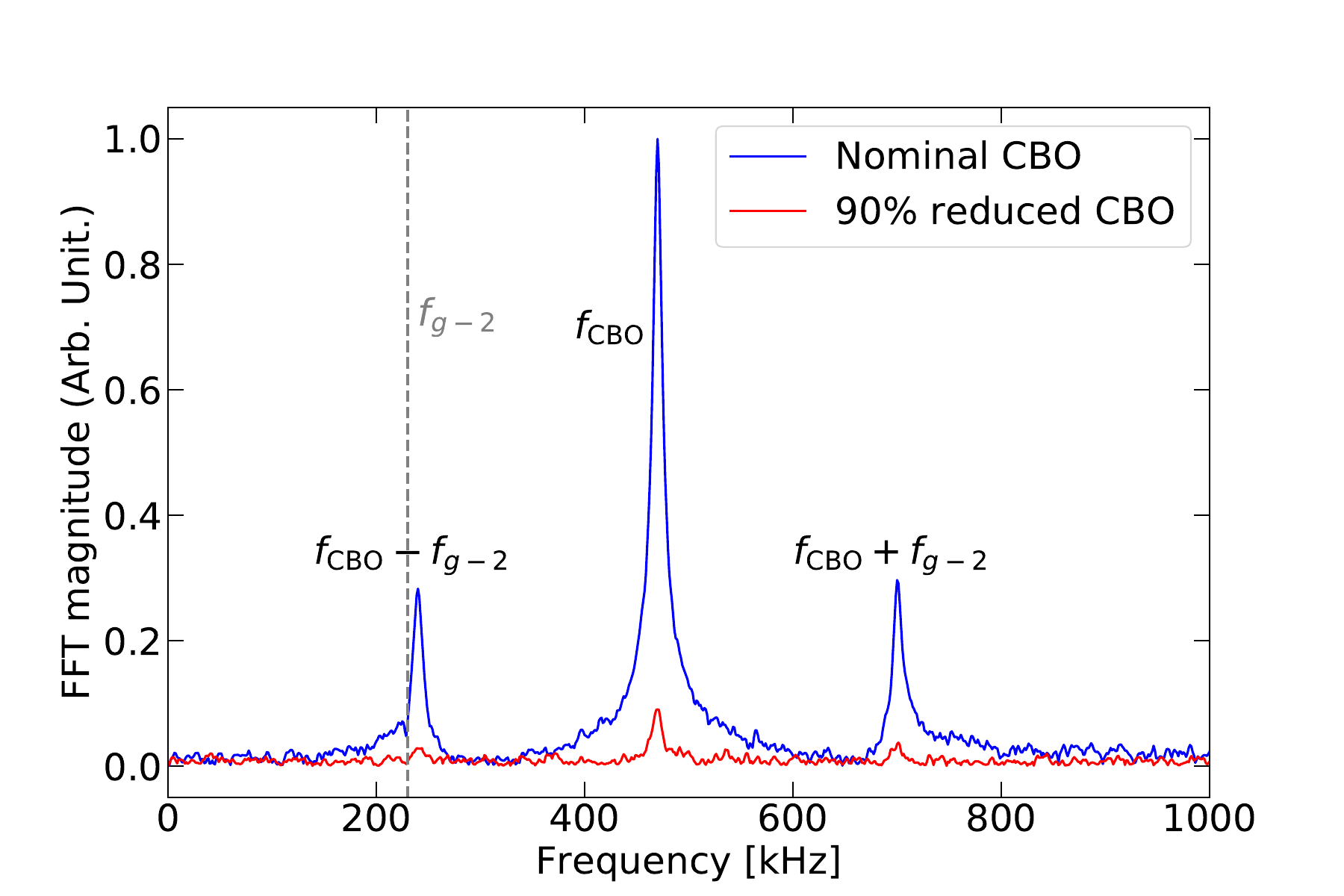}
	\caption{FFT spectrum of the residuals from fitting Monte Carlo simulation   data to the five-parameter function. $N_0$, $A$ and $\phi$ are modulated at  $f_\text{CBO}$ in the simulation. With $f_\text{CBO}\approx 2f_{a}$, the peak at   $f_\text{CBO}-f_{a}$ may partially overlap with the $f_{a}$ signal. Details on the estimation of the modulation parameters are given in Ref. \cite{ref:bnl_final_report}.} 
	\label{fig:fcbo_fft}
\end{figure}

When the narrow time bunch muon beam enters the storage ring, each calorimeter sees the beam moving in and out with this CBO frequency. Because the detector acceptance depends on the radial position of the muon decay, this inward and outward motion of the bunched beam amplitude modulates the positron decay time spectrum. The principal issue is that its lower side band may overlap with $f_\text{a}$ if  $f_\text{CBO}$ is close to $2f_\text{a}$, thereby pulling the $g-2$ phase. The effect can be significantly reduced, perhaps eliminated, if the CBO can be suppressed by an order of magnitude (see Figure \ref{fig:fcbo_fft}). Note that this effect does not appear at the frequencies related to the vertical CBO. Therefore, our main focus will be on the horizontal motion.

\begin{figure}[h!]
	\centering
	\includegraphics[width=0.9\linewidth]{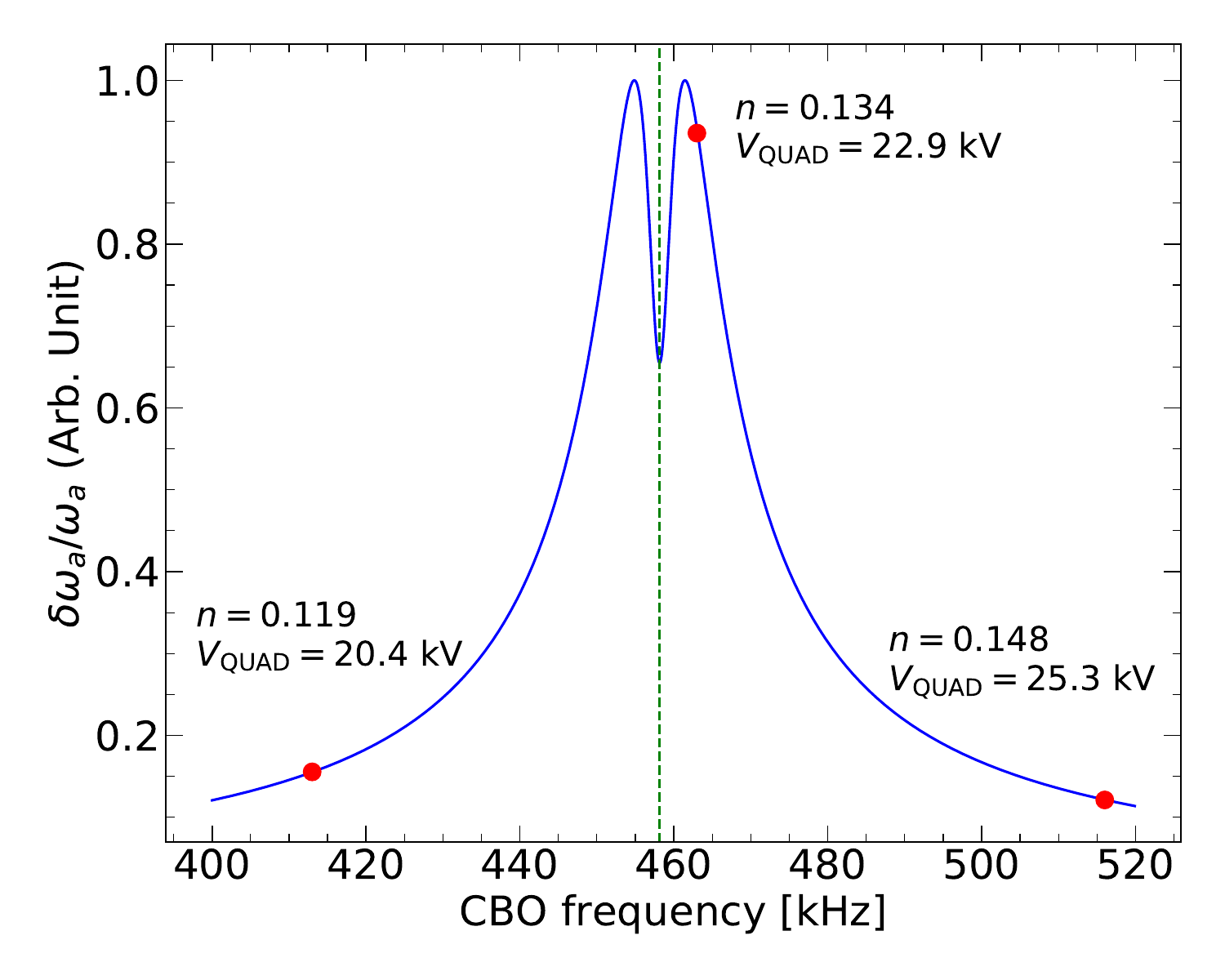}
	\caption{The maximum possible relative pull ($\delta \omega_\text{a}/\omega_\text{a}$) versus the CBO   modulation frequency, if not addressed by the fitting function. The   systematic error $\delta \omega_\text{a}/\omega_\text{a}$ has a strong dependence on  how   close $f_\text{CBO}$ is to $2f_\text{a}$. The estimations are based on simulations with varying field index values $n$. The red points are selected arbitrarily to show the $n$-values and the quadrupole voltages corresponding to various CBO frequencies.} 
	\label{fig:fcbo_with_n}
\end{figure}

The CBO frequency depends on the $n$-value of the ring, {\it viz.} the quadrupole voltage. As the CBO frequency approaches $2f_\text{a}$, the systematic pull on $\omega_\text{a}$ becomes significant, as discussed in section 5.1 of Ref.~\cite{ref:bennett_2007}. Figure~\ref{fig:fcbo_with_n} shows the relative systematic error from the CBO as a function of $f_\text{CBO}$ caused by the lower
sideband overlapping with $f_\text{a}$. An obvious strategy to minimize this issue would be to change the $n$-value by increasing or decreasing the quadrupole high voltage in order to move away from the central region of Figure \ref{fig:fcbo_with_n}.  However, raising the high voltage (higher $n$) increases the possibility of the quadrupoles sparking, and lowering the voltage (lower $n$) decreases the acceptance of the storage ring, resulting in fewer muons stored.

Coherent betatron oscillations were a well-known effect in the BNL data, which was discussed in the E821 papers~\cite{ref:bnl_final_report,ref:cbo_analysis_1,ref:bennett_2007}. In the analysis of Run 99 and later, Equation \ref{eq:five_param} is modulated with an additional term to modulate the acceptance, phase and asymmetry:
\begin{equation}
C(t)= 1-e^{-t/\tau_\text{CBO}} A_\text{CBO} \cos(2 \pi f_\text{CBO}t + \phi_\text{CBO}),
\label{eq:five_param_with_cbo}
\end{equation}
where $\tau_\text{CBO}$ is the lifetime of the CBO, $A_\text{CBO}$ is the modification of the acceptance, and $\phi_\text{CBO}$ is the CBO phase. This additional parametrization resulted in an acceptable $\chi^2/$dof, and a conservative systematic error of $\pm 0.07$ ppm (70 ppb) for this effect was assigned \cite{ref:bnl_final_report}. Because the total E989 error budget for the muon frequency $\omega_\text{a}$ is four times smaller, we need to reduce the CBO contribution on $\omega_\text{a}$ from 70 ppb to 30 ppb. Reducing the CBO amplitude will significantly reduce the risk associated with the choice of the functional form describing it.

The muon populations at the edge of the beam tend to hit the walls of the vacuum chamber before decaying, which causes a modified decay rate measurement at the detectors. This systematic error was reduced to 90 ppb level in the E821 experiment. It was achieved by scraping the beam halo against circular collimators with a 9~cm diameter through the asymmetrical powering method.

Immediately before the muon beam injection, the top and bottom plates on quadrupoles 1-4 were powered asymmetrically, as were the vertical side plates on quadrupoles 2 and 4. This asymmetric powering  shifted the beam both vertically and horizontally onto the collimators. Five to fifteen microseconds later, the plate voltages were returned to a symmetric configuration with a 5~$\mu$s time constant, centering the beam in the storage region ~\cite{ref:bnl_final_report,ref:bnl_quad}. This procedure removed particles near the aperture and significantly reduced muon losses from the storage ring during the data collection period. The asymmetrical powering method has been utilized in the first two runs of E989 as well. An approximately $ 7~\mu$s scraping time  was used~\cite{ref:fnal_quad}. 

A major issue with this scraping method is that the beam can move through a resonance while going from the scraping high-voltage  configuration to the final symmetric high voltage, causing unwanted muon losses. 

In the next section, we introduce the RF reduction method for damping the CBO. Application of RF fields allows us to scrape the beam as well, while avoiding the unwanted muon loss. It is worth emphasizing that reducing the CBO effect ensures a better fitting result, in other words, a more sensitive $a_\mu$ measurement.

\section{The RF reduction method: Analytical calculation of CBO reduction} {\label{Yuri}}

The manipulation of charged particle beams in accelerators using radio frequency (RF) techniques has a long history, and many electron, proton and heavy ion storage rings have been built and operated.  RF beam separators have also been utilized to improve the purity of secondary beams. Our method relies on damping the coherent betatron oscillations by applying a transverse RF electric field at the CBO frequency. Because of the limited muon lifetime, the application of the RF field should be finished within a few tens of microseconds. 

Reduction of coherent betatron oscillations by RF fields was first proposed in 2003 as a potential upgrade to the BNL E821 storage ring experiment~\cite{Orlov:2003}.  We present that discussion here, since this reference is not published. 

To damp the CBO, a harmonically varying horizontal dipole electric field $E_x(t)$ is applied to the beam {\it out of phase} with the CBO. The RF field is applied to the two  vertical quadrupole plates, which begin at the longitudinal position $s_0$ and extend to  $s = s_0 + \ell$.
\begin{equation}
E_x(t)=E_\text{x0}f(s)\cos[\omega_\text{CBO}t + \theta_0],
\end{equation}
where, $\theta_0$ is the initial CBO phase and $f(s)$ is defined as
\[
f(s)= 
\begin{cases}
1 & s_0<s<s_0+\ell\\
0 & \text{otherwise}.
\end{cases}
\]
Setting the muon injection time to be $t=0$,  we obtain Equation \ref{eq:eq_of_mot_x}, Hill's equation for the horizontal component, with a harmonic driving term:
\begin{equation}
\ddot x + \omega_\text{c}^2(1-n)x=\frac{\omega_\text{c}  E_\text{x0}}{B_0}f(t)\cos[\omega_\text{CBO}t +\theta_0], 
\label{eq:perturb_eq_of_motion}
\end{equation}
where
\begin{equation}
f(t)= 
\begin{cases}
1 & Tj<t<Tj+l/v, ~~j=0,1,2,...,N\\
0               & \text{otherwise}.
\end{cases}
\label{eq:f_t}
\end{equation}
The definition of $f(t)$ means that the muon passes through the damping electric field $E_x(t)$ periodically with the revolution period $T=2\pi/\omega_\text{c}$, $N+1$ times, after which the RF perturbation is turned off. The exact solution of Equation \ref{eq:perturb_eq_of_motion} with $f(t)$ given by Equation \ref{eq:f_t} is  
\begin{equation} 
x(t)=a(t)e^{i\omega_\text{x} t}+ a^*(t)e^{-i\omega_\text{x} t},
\end{equation}
\begin{equation}
a=a_0-\frac{iE_\text{x0}/B_0}{2\sqrt{1-n}}\int dt f(t)e^{-i\omega_\text{x} t}\cos[\omega_\text{CBO}t +\theta_0], 
\label{eq:a_0}
\end{equation}
where $a_0$ corresponds to $t=0$. From Equation \ref{eq:a_0}, one gets the solution at time $t=(N+1)T+l/v$, the time that the CBO damping finishes. At this time
\begin{equation}
\begin{split}
a&=a_0-i e^{-i\theta_0} \frac{N+1}{4 \sqrt{1-n}}\left(\frac{l E_\text{x0}}{ v B_0}\right) \\ 
&\times \left[ 1+\frac{e^{2i\theta_0}}{N+1} \frac{1-e^{-2i\omega_{x} (N+1) T}}{1-e^{-2i\omega_{x}T}} \right].
\end{split}
\label{eq:yuri_final}
\end{equation}
Ideally, $a \rightarrow 0$, or very close to zero after $N+1$ turns. After discussing the simulations of the RF method, we will compare this analytical formula with a simulation result in Section \ref{sect:single_part_sim}.

\section{Simulations of the RF reduction method}{\label {sect:simulation}}

\begin{figure}[h!]
	\centering
	\includegraphics[width=0.5\linewidth]{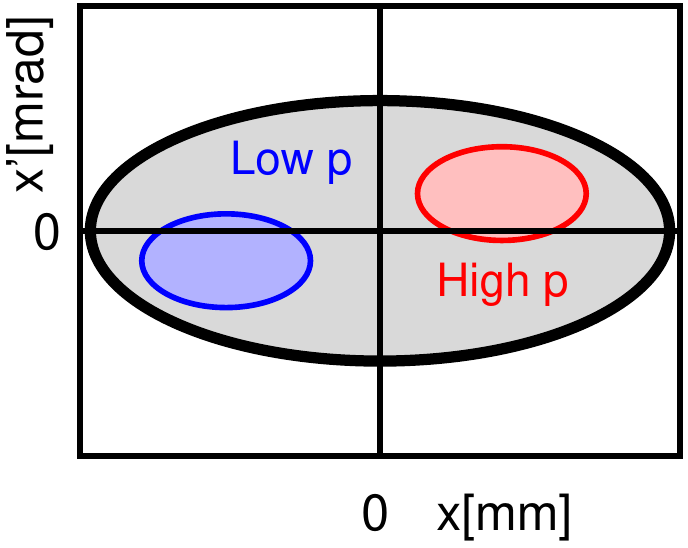}
	\caption{Illustration of the radial phase space of the ring. The black ellipse represents the ring acceptance; the red and blue ellipses represent the high- and the low-momentum populations, respectively, with respect to the design (magic) momentum.}
	\label{fig:rf-cartoon}
\end{figure}

In this and subsequent sections, we follow the conventions of Figure \ref{fig:rf-cartoon} for both phase space ellipses and individual particles of low- and high- momentum in phase space.

\subsection{Simulation strategy}

\begin{figure}[h]
	\centering
	\begin{subfigure}[h]{\linewidth}
		\centering
		\includegraphics[width=0.8\textwidth]{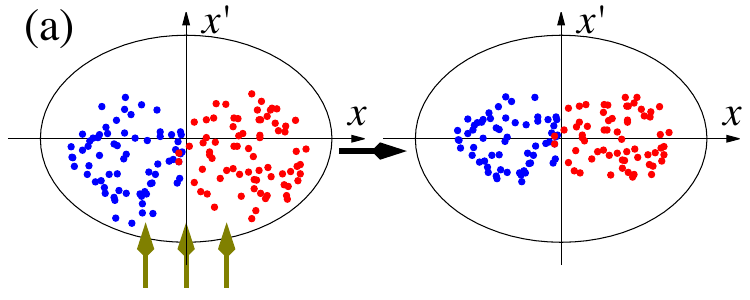}
		\label{fig:dipRFcentering}
	\end{subfigure}
	~
	\begin{subfigure}[h]{\linewidth}
		\includegraphics[width=\linewidth]{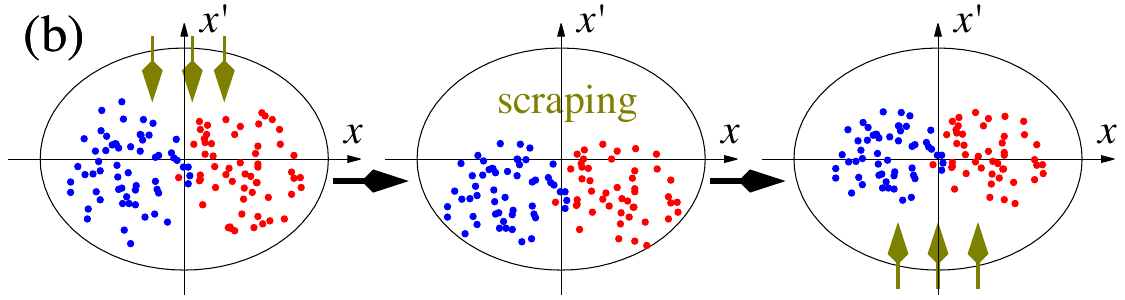}
		\label{fig:dipRFscraping}
	\end{subfigure}
	\caption{Illustration of the two different uses of the RF dipole fields: (a) CB reduction and (b) scraping. The black elliptic arcs are the transverse phase space acceptances, the red and blue spots are the high- and low-momentum populations, respectively, and the green arrows are the force exerted by the RF field.}
	\label{fig:dipRFmodes}
\end{figure}

A dipole RF field with the resonant frequency can move the centroid of the beam along the vertical axis of the transverse phase space. It moves the high- ($\delta p/p_\text{m}>0$) and low- ($\delta p/p_\text{m}<0$) momentum populations together when they have the same CBO phase.  Figure \ref{fig:dipRFmodes} shows the effect of the dipole RF field with $f_\text{CBO}$ on the phase space. Figure \ref{fig:dipRFmodes} (a) shows that by a correct choice of the RF phase, the beam can be shifted on the vertical axis to reduce the CBO amplitude. Figure \ref{fig:dipRFmodes} (b) shows that the dipole RF field can also be used in "scraping mode" before the measurement period by switching to the opposite RF phase.

\begin{figure}[h]
	\centering
	\includegraphics[width=0.8 \linewidth]{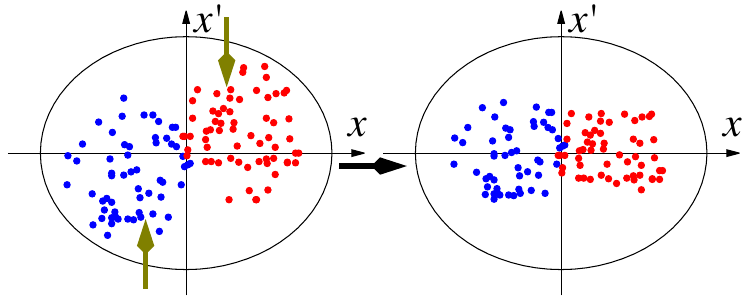}
	\caption{Illustration of the beam phase space in the presence of a RF quadrupole field. This is applicable only to the horizontal phase space. The red and blue spots are high- and	low-momentum muon populations, respectively, and the green arrows indicate the electric force on the beam.} 
	\label{fig:quadRFsqueezing}
\end{figure}

In contrast to the RF dipole mode, the RF quadrupole mode moves the high- and low-momentum populations in opposite directions. The beam width decreases as they reach the design orbit (see Figure \ref{fig:quadRFsqueezing}).

\subsection{Simulation conditions}
The simulations of the RF reduction were done with a precision tracking simulation tool written in \texttt{C++}~\cite{ref:tracking_benchmarks}. All of the essential ring elements were implemented in the program, such as
the electrostatic quadrupoles~\cite{ref:bnl_quad}, the fast kicker magnets~\cite{Schreckenberger:2018njd}, the vacuum chamber, and the collimators. In these simulations, a 3~kV RF voltage was applied to one 3.2~m azimuthal quadrupole section.

As described above, the beam enters the E989 muon storage ring through the inflector section (shown in Figure \ref{fig:ring_from_top}), after which the simulations start with a beam distribution that was previously obtained by independent Monte Carlo simulations \cite{ref:dRubin}. After the inflector, the fast kicker applies a vertical pulsed magnetic field to move the beam towards the design orbit. The beam motion in the phase space is depicted in Figure \ref{fig:kicker_at_phase_space}. 

The electric quadrupole plates have a cross section of $10 \times 10$~cm$^2$~\cite{ref:bnl_quad}. An RF voltage of 0.3 kV amplitude is applied to all eight quadrupole sections, covering 43\% of the ring circumference.  Particles are stored for up to 200~$\mu s$ in the simulations. 

\subsection{Single Particle Simulations} {\label{sect:single_part_sim}}

As a proof of principle, single particle simulations were conducted using three particles of different momenta: magic ($\delta p/p_\text{m}=0$), high- ($\delta p/p_\text{m}=+0.25\%$) and low- ($\delta p/p_\text{m}=-0.25\%$) momentum muons. In the absence of an RF field, each particle follows an elliptical trajectory in phase space, whose origin and radius are determined by its momentum and initial position. 

\begin{figure}
	\centering
	\begin{subfigure}{0.9\linewidth}
		\includegraphics[width=\textwidth]{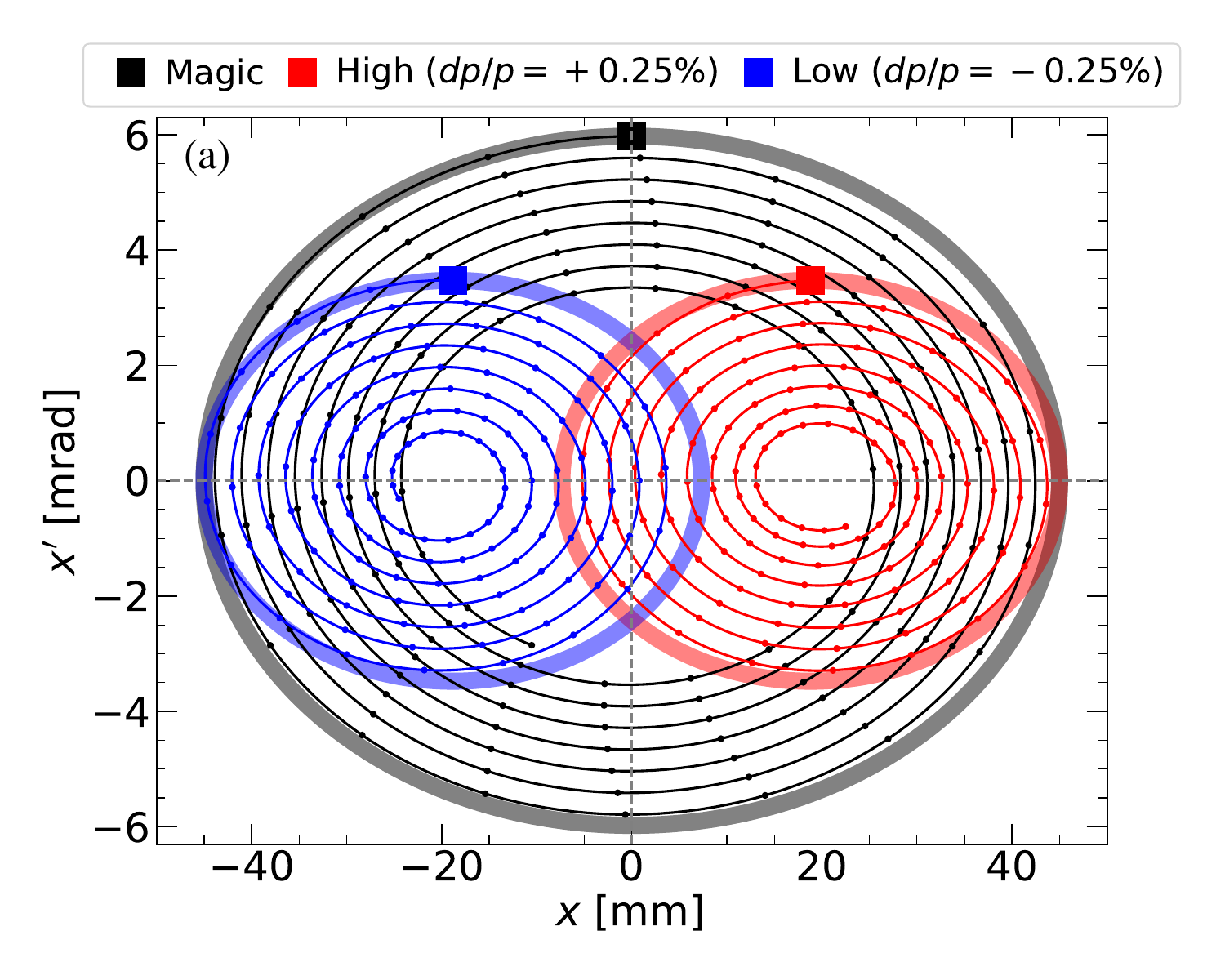}
		\label{fig:dipoleRFinphase_a}
	\end{subfigure}
	~
	\begin{subfigure}{0.9\linewidth}
		\includegraphics[width=\textwidth]{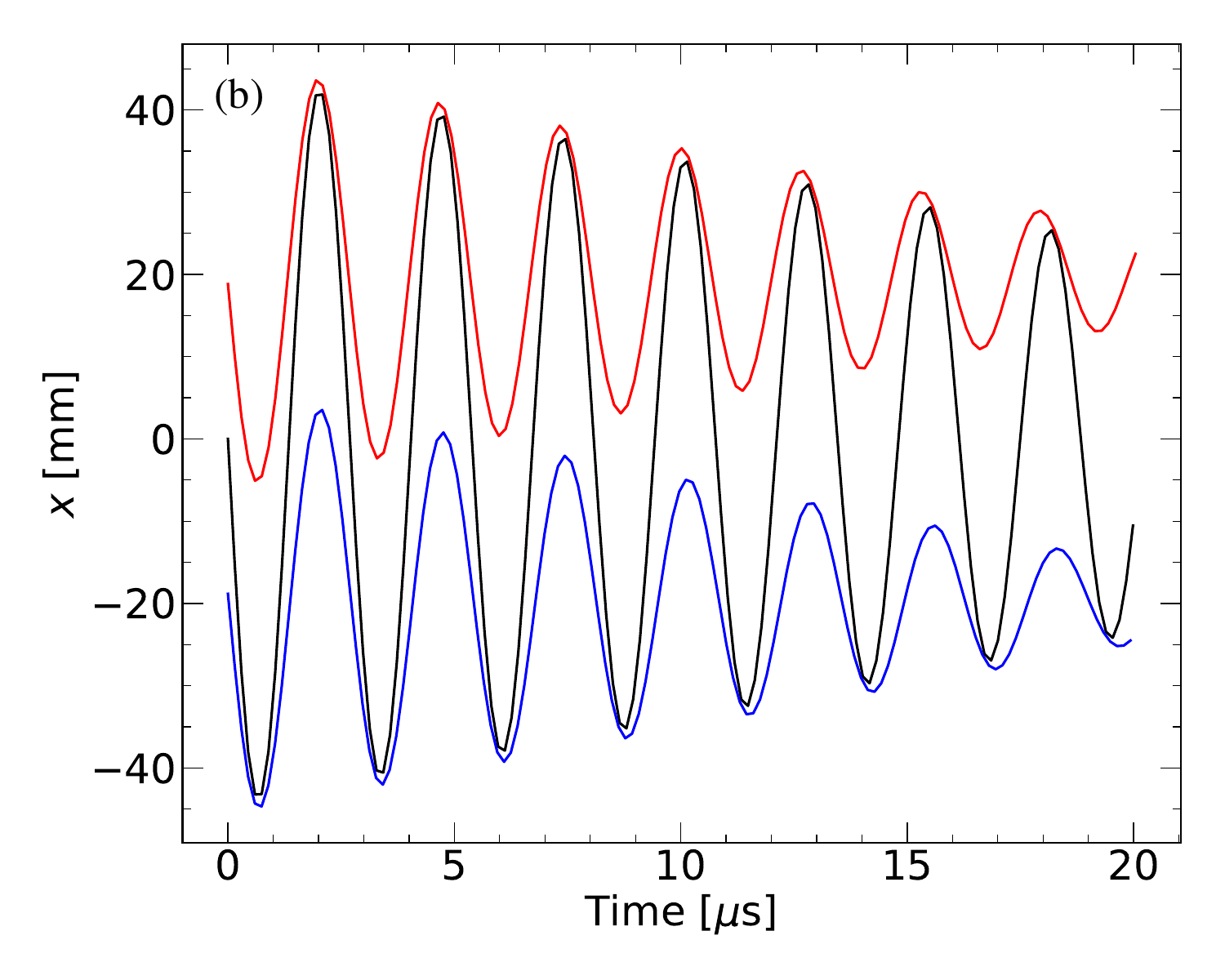}
		\label{fig:dipoleRFinphase_b}
	\end{subfigure}
	\caption{The horizontal phase space and position as a function of time for three muons with an applied RF dipole field. The simulations were made with particles of three momenta; black: $\delta p/p_\text{m} = 0$; red: $\delta p/p_\text{m} =+ 0.25\%$; blue: $\delta p/p_\text{m} =- 0.25\%$. The dipole RF field is applied in phase with the muon phase. (a): Horizontal phase space trajectories for the three momenta when an RF dipole field is applied. The thick ellipses show the phase-space trajectories without RF. The filled squares show the initial phase space value for each of the ellipses at the beginning of the simulation. (b): Horizontal position $x$ as a function of time for this simulation.}
	\label{fig:dipoleRFinphase}
\end{figure}

\begin{figure}
	\centering
	\begin{subfigure}{0.9\linewidth}
		\includegraphics[width=\textwidth]{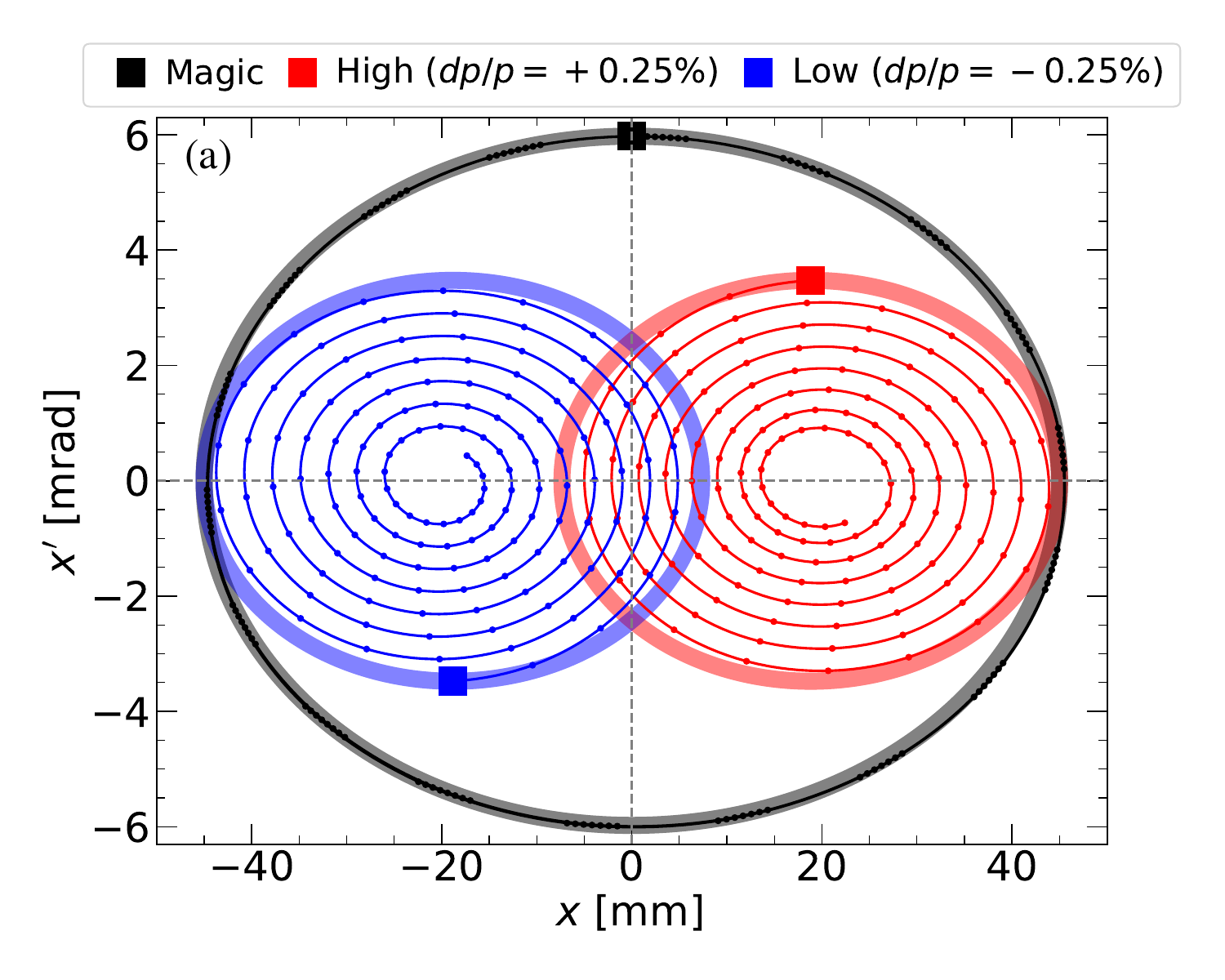} 
		\label{fig:quadrupoleRFinphase_a}
	\end{subfigure}
	~
	\begin{subfigure}{0.9\linewidth}
		\includegraphics[width=\textwidth]{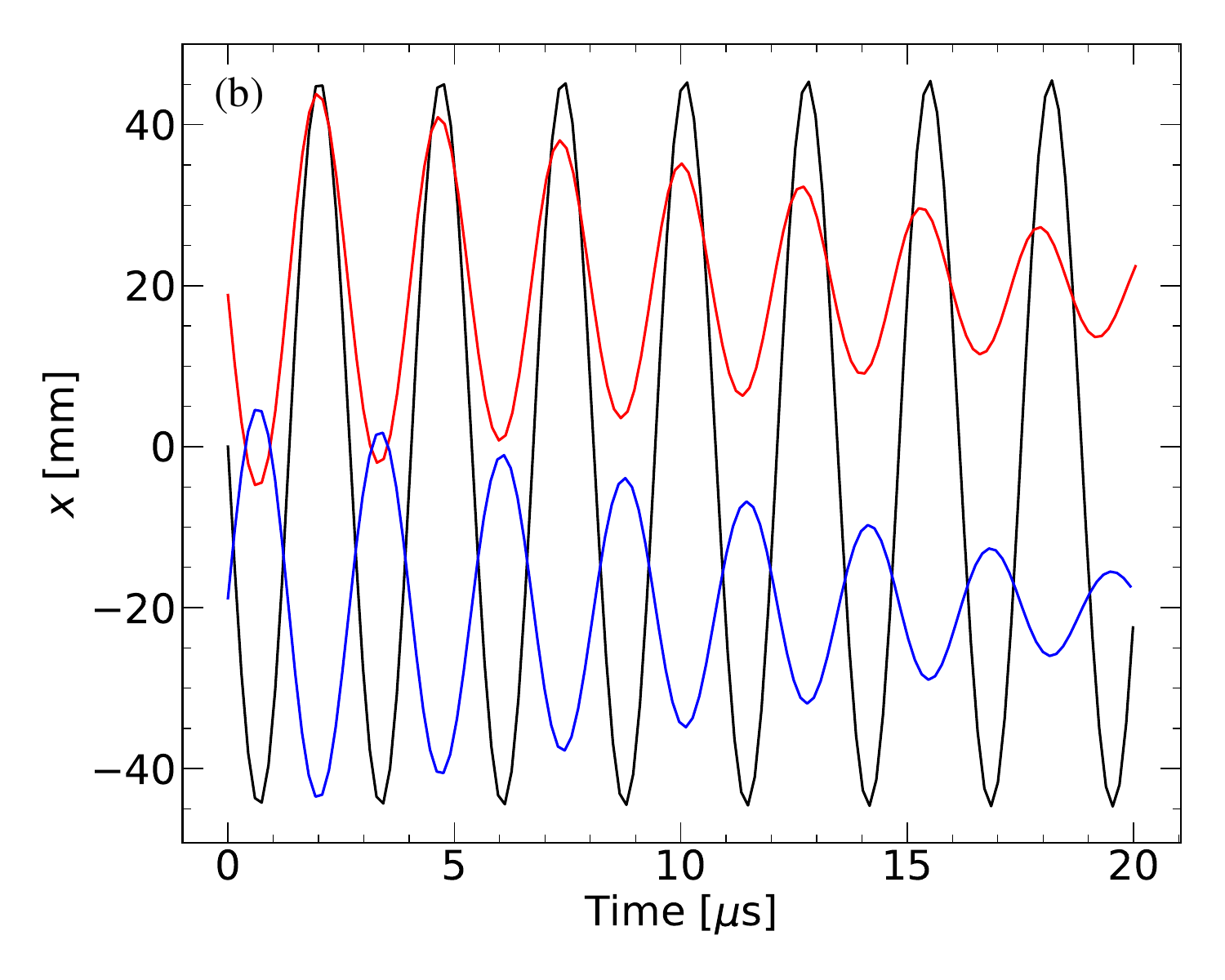}
		\label{fig:quadrupoleRFinphase_b}
	\end{subfigure}
	\caption{The horizontal phase space and position as a function of time when an RF quadrupole was applied for three momenta in the simulations; black: $\delta p/p_\text{m} = 0$; red: $\delta p/p_\text{m} =+ 0.25\%$; blue: $\delta p/p_\text{m} =- 0.25\%$. The magic momentum muons were not affected. (a): The horizontal phase space trajectory when an RF quadrupole field was applied. The filled squares show the initial phase space value for each of the trajectories. Note that the initial phases of the high- and low- momentum muons are opposite. (b): Horizontal position $x$ as a function of time during the application of an RF quadrupole.}
	\label{fig:quadrupoleRFinphase}
\end{figure}

Figure \ref{fig:dipoleRFinphase} shows the phase space trajectory of each particle in the presence of the dipole RF field applied with the optimum phase. The method works best if the particles are in phase. If they are out of phase, then a dipole RF increases the amplitude of one, and shrinks the amplitude of the other. On the other hand, the RF quadrupole mode shrinks the high- and low-momentum particle orbits if they are out of phase (Figure \ref{fig:quadrupoleRFinphase}). Note that the particle with magic momentum is not affected by the RF quadrupole field as it oscillates around the design orbit. 

\begin{figure}[h!]
	\centering
	\includegraphics[width=0.9\linewidth]{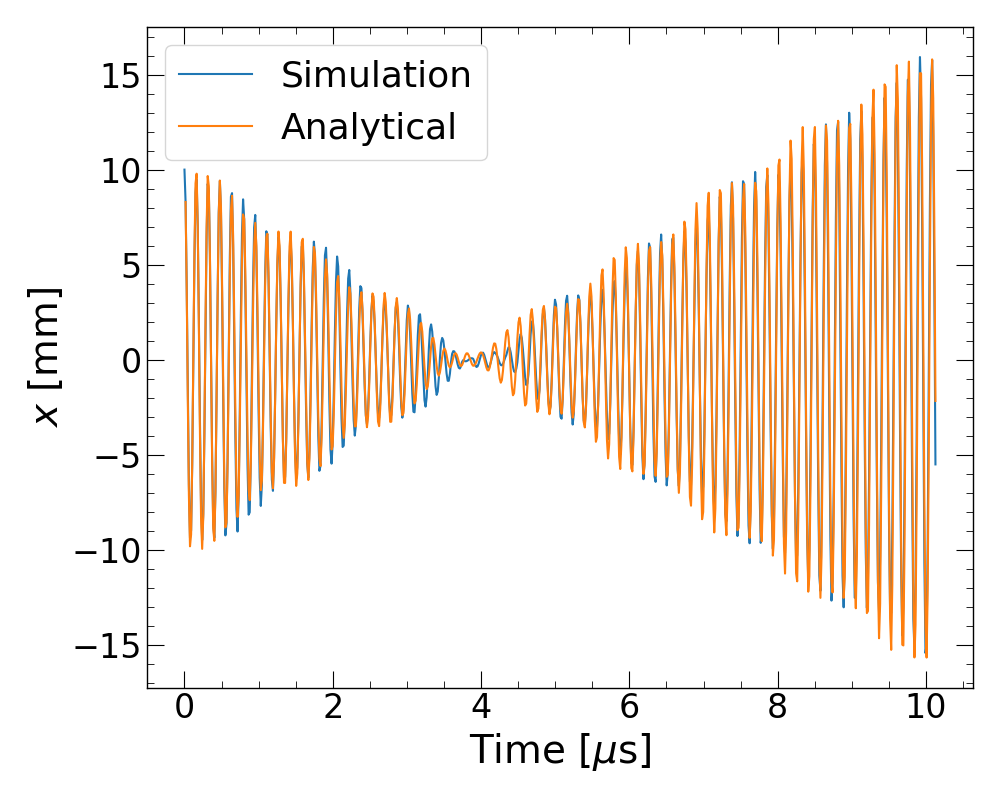}
	\caption{Comparison of the analytical and simulation results for reducing the CBO using the parameter values in the text. Application of the RF field should stop after reaching the minimum amplitude ($\approx 4~\mu$s).} 
	\label{fig:YuriFormulaConfirm}
\end{figure}

Figure\ref{fig:YuriFormulaConfirm} shows the excellent agreement between the analytical calculation presented in Section \ref{Yuri}  and the result  from a single particle simulation. The parameters used in the calculation are:  $n=0.12$, $l=3.2$ m, $E_\text{x0}=100~\text{kV/m}$, $B_0=1.45$ T, $f_\text{c}=6.71$ MHz, $f_\text{CBO}=400$ kHz  and $\theta_0=134^\circ$.  Note that the
RF field should be turned off at the minimum, after which the CBO starts growing because of resonance. In this example, it takes roughly $N=23$ oscillations to minimize the CBO amplitude.  As seen here, amplitude damping with an opposing RF kick is the only effect for the single particle case. However, coupling between high- and low-momentum populations brings about a phase shift, a more dominant reduction effect. This will be presented in the next section.

\subsection{Multiparticle Simulations}

The multiparticle simulations were done with roughly 30,000 muons entering the storage ring. 95\% of the particles hit the vacuum chamber or collimators and were lost after several turns around the ring. 

The RF phase was initially optimized for the maximum CBO reduction without applying any scraping. After determining the optimum phase, the dipole RF field was applied from 2 to 7 $\mu$s for the scraping and from 10 to 20 $\mu$s for CBO reduction. Note that these two steps should have opposite phases.

\begin{figure}[h!]
	\centering
	\includegraphics[width=\linewidth]{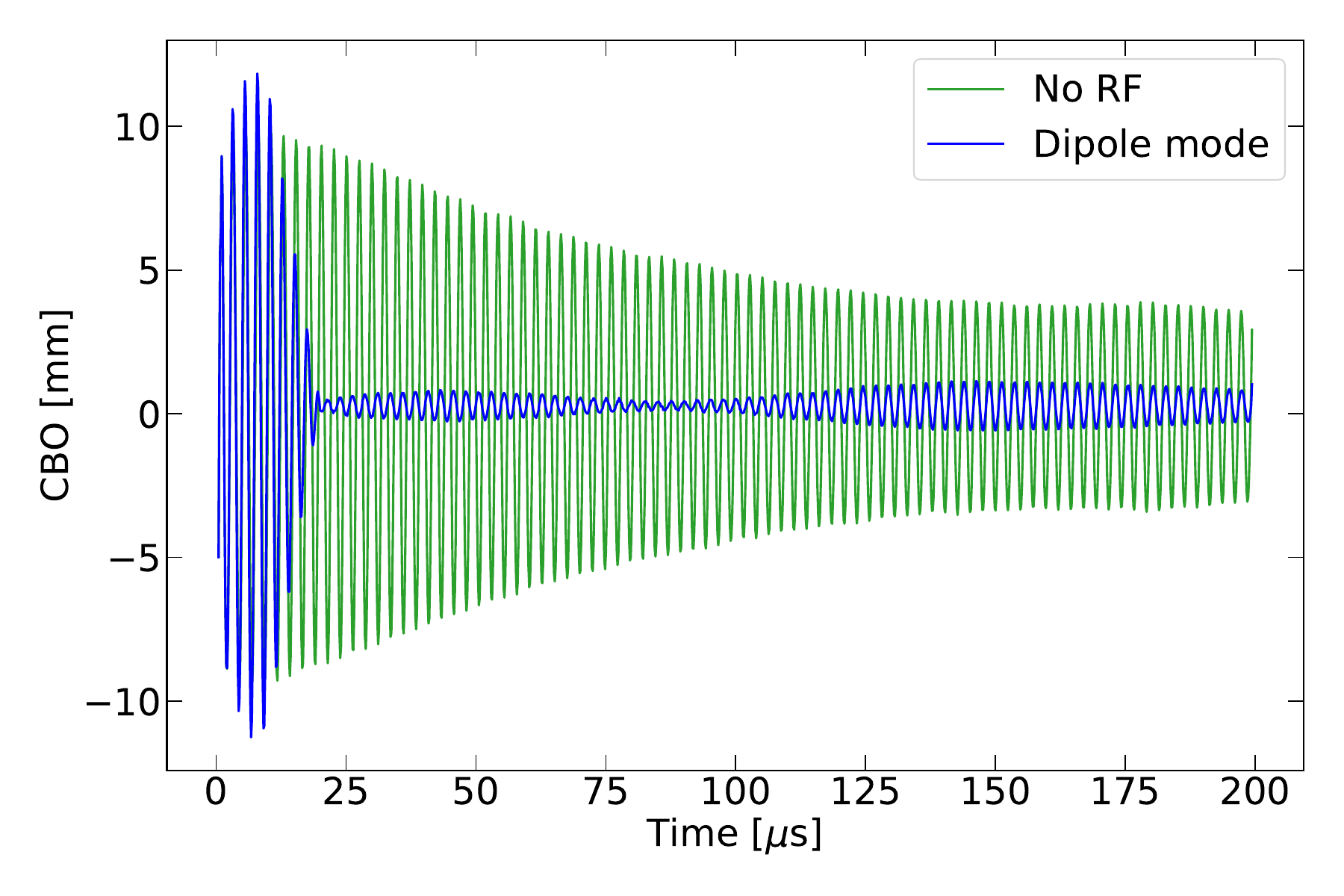}
	\caption{Beam centroid versus time. The simulations were made with (blue) and without (green) the dipole RF field.}
	\label{fig:CBOdipole}
\end{figure}

\begin{figure}[h!]
	\centering
	\includegraphics[width=\linewidth]{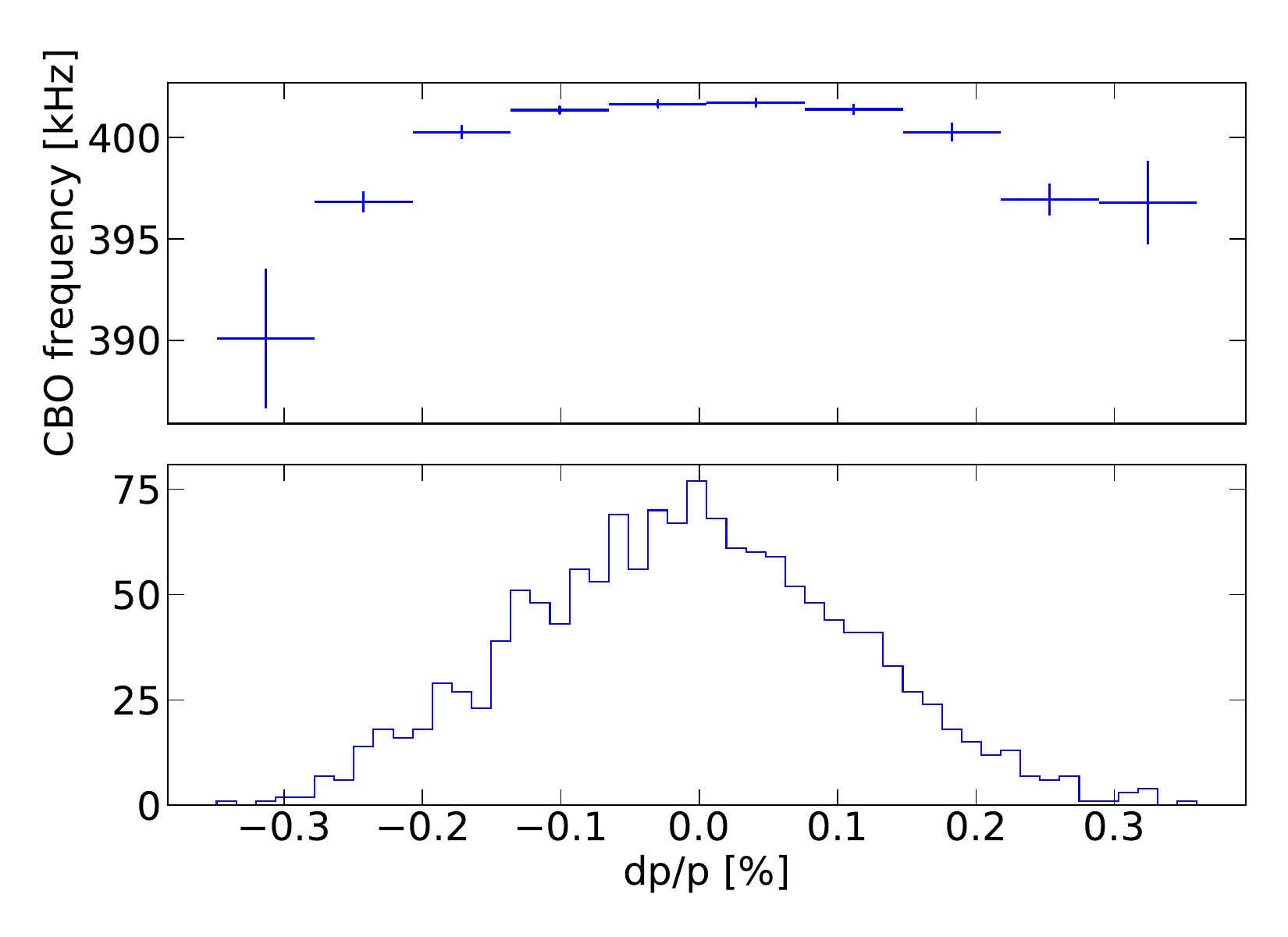}
	\caption{CBO frequency of a particle has a momentum dependence, which is known as chromaticity. The histogram in (a) was obtained by multiparticle simulations, where the particles had a momentum distribution as shown in (b). Because of this variation, the CBO of the beam gets modulated after the RF field is turned off (Figure \ref{fig:CBOdipole}.) }
	\label{fig:cbofreqvsmomentum}
\end{figure}

Figure \ref{fig:CBOdipole} compares the CBO without RF, and with the applied RF field at the optimum phase. The CBO decoherence time is $100-200~\mu$s in the absence of an RF field. The damping time depends on the momentum spread of the beam and the multipole components of the electric focusing field. Because of the electric
20-pole component, the CBO frequency of the different momentum muons varies, eventually leading to a decoherence (See Figure \ref{fig:cbofreqvsmomentum}). However, the application of the RF field before 20 $\mu$s improves the CBO damping by almost an order of magnitude. The beating after 20 $\mu$s originates from the frequency spread of the beam shown in Figure \ref{fig:cbofreqvsmomentum}.

\begin{figure}
	\centering
	\includegraphics[width=\linewidth]{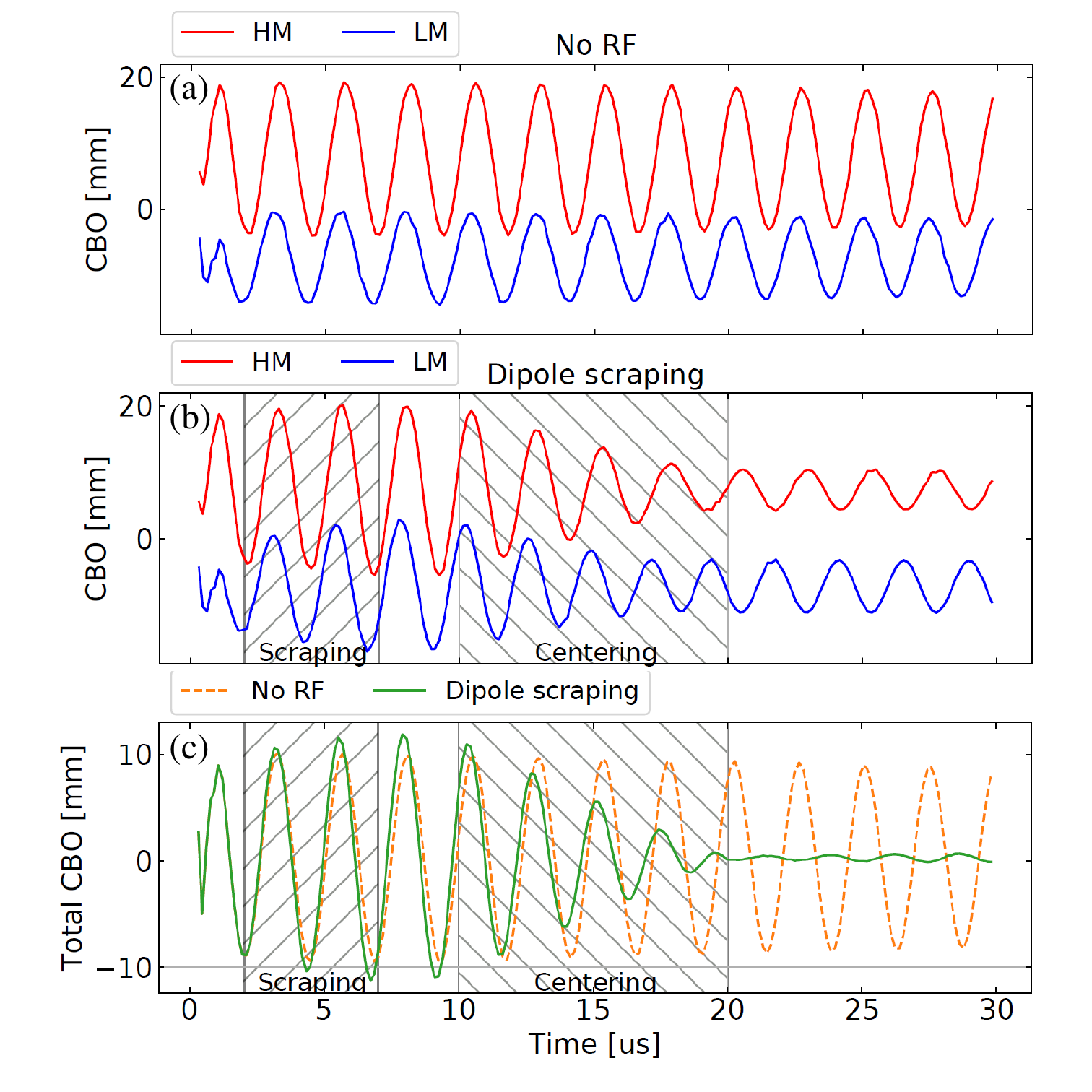}
	\caption{The CBO of the high (red), low (blue) and combined (green) momentum as obtained in the simulations. (a) No RF was applied. (b) RF dipole field damps the CBO amplitude and shifts the phase of both populations. (c) The combined CBO before and after the application of the RF dipole.  The dashed orange line is without the application of RF.} 
	\label{fig:CBOdipole_separate}
\end{figure}

The mechanism of the CBO reduction by a dipole RF field is shown in Figure \ref{fig:CBOdipole_separate}. The initial conditions were obtained from Monte Carlo simulations \cite{ref:dRubin}. As Figure \ref{fig:CBOdipole_separate} (b)  shows, the RF dipole field reduces the CBO amplitude of both the high- and low-momentum populations.  However, it is worth emphasizing that the reduction is not limited to a simple damping effect. In fact, the dominant effect is the phase shift between the two populations. The CBO phase of the high-momentum population advances with the RF kick, since it is closer to the origin of the phase space (see the tilt of the beam at Figure \ref{fig:kicker_at_phase_space}). 

\begin{figure}
	\centering
	\includegraphics[width=\linewidth]{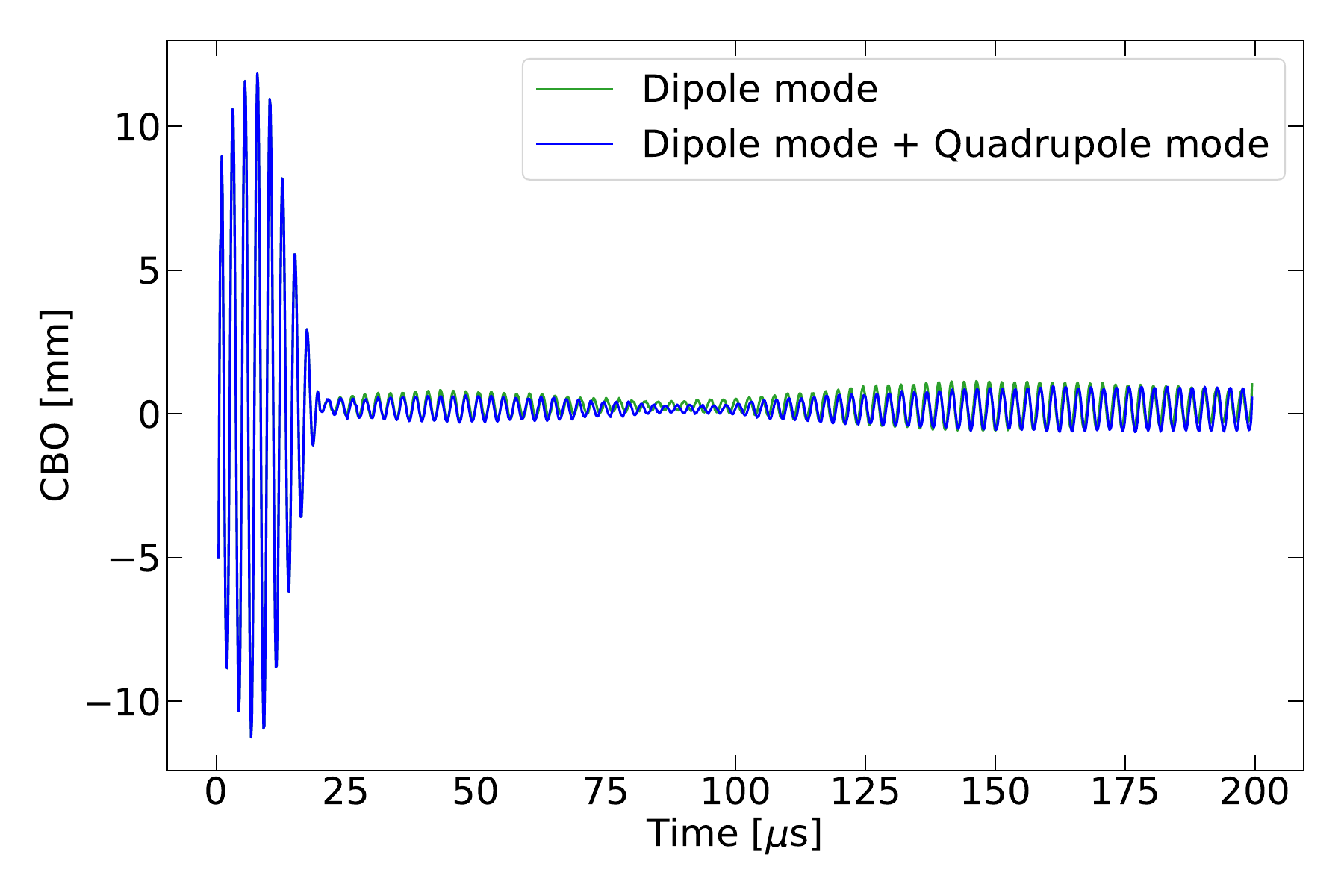}
	\caption{The beam centroid versus time. The simulations were made with dipole RF (green) and with both dipole and quadrupole RF (blue). The difference in centroid is negligible. } 
	\label{fig:CBOquadrupole}
\end{figure}

\begin{figure}
	\centering
	\includegraphics[width=\linewidth]{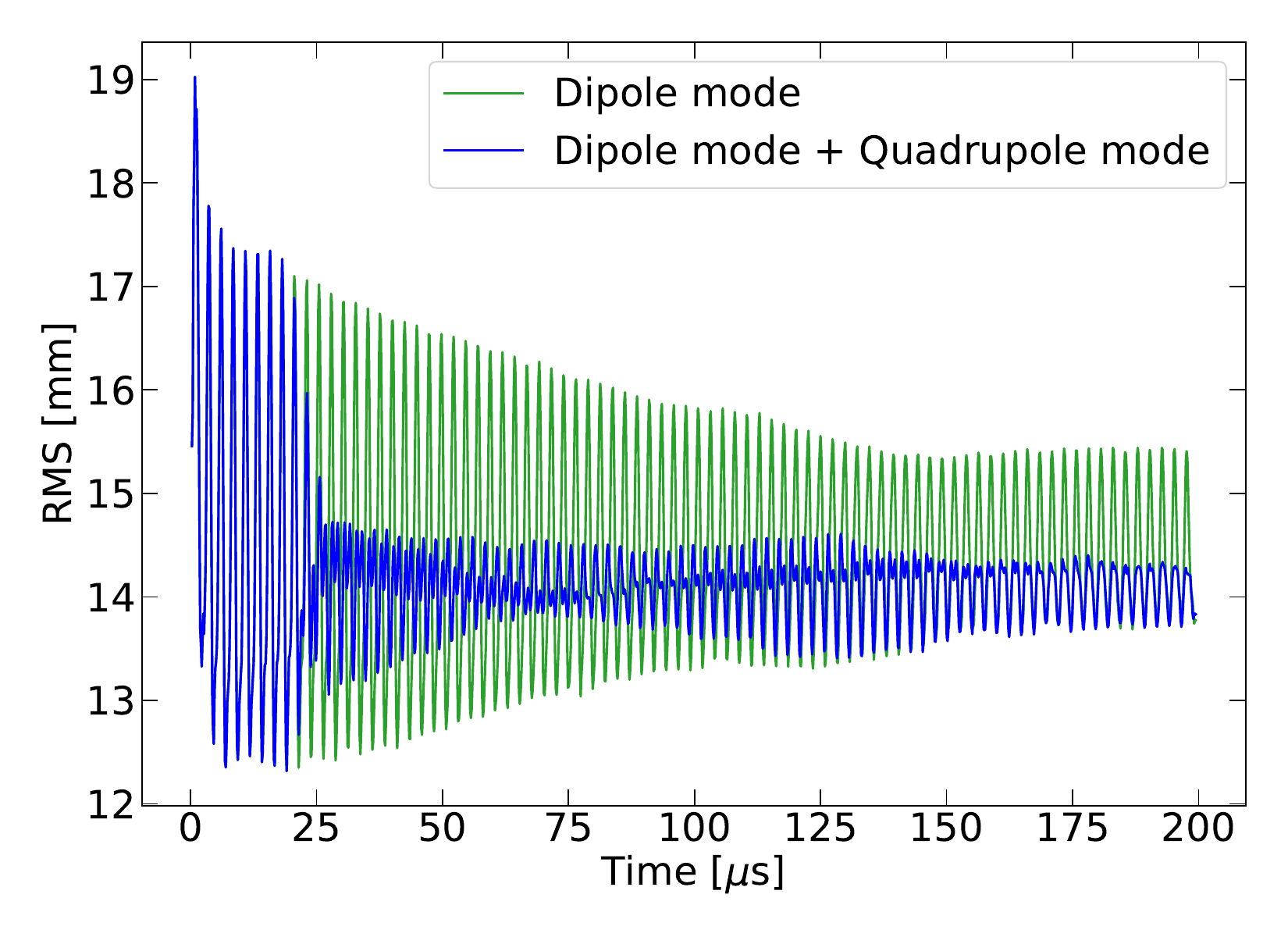}
	\caption{Beam RMS versus time. The simulations were made with (blue) and without (green) the quadrupole RF field.}
	\label{fig:RMSquadrupole}
	
\end{figure}

Applying the RF quadrupole field after the dipole mode does not have any effect on the beam centroid, as shown in Figure \ref{fig:CBOquadrupole}. However the RMS modulation of the beam is significantly reduced by the application of the RF field, as shown in Figure \ref{fig:RMSquadrupole}.  

\begin{figure}
	\centering
	\begin{subfigure}{\linewidth}
		\includegraphics[width=\textwidth]{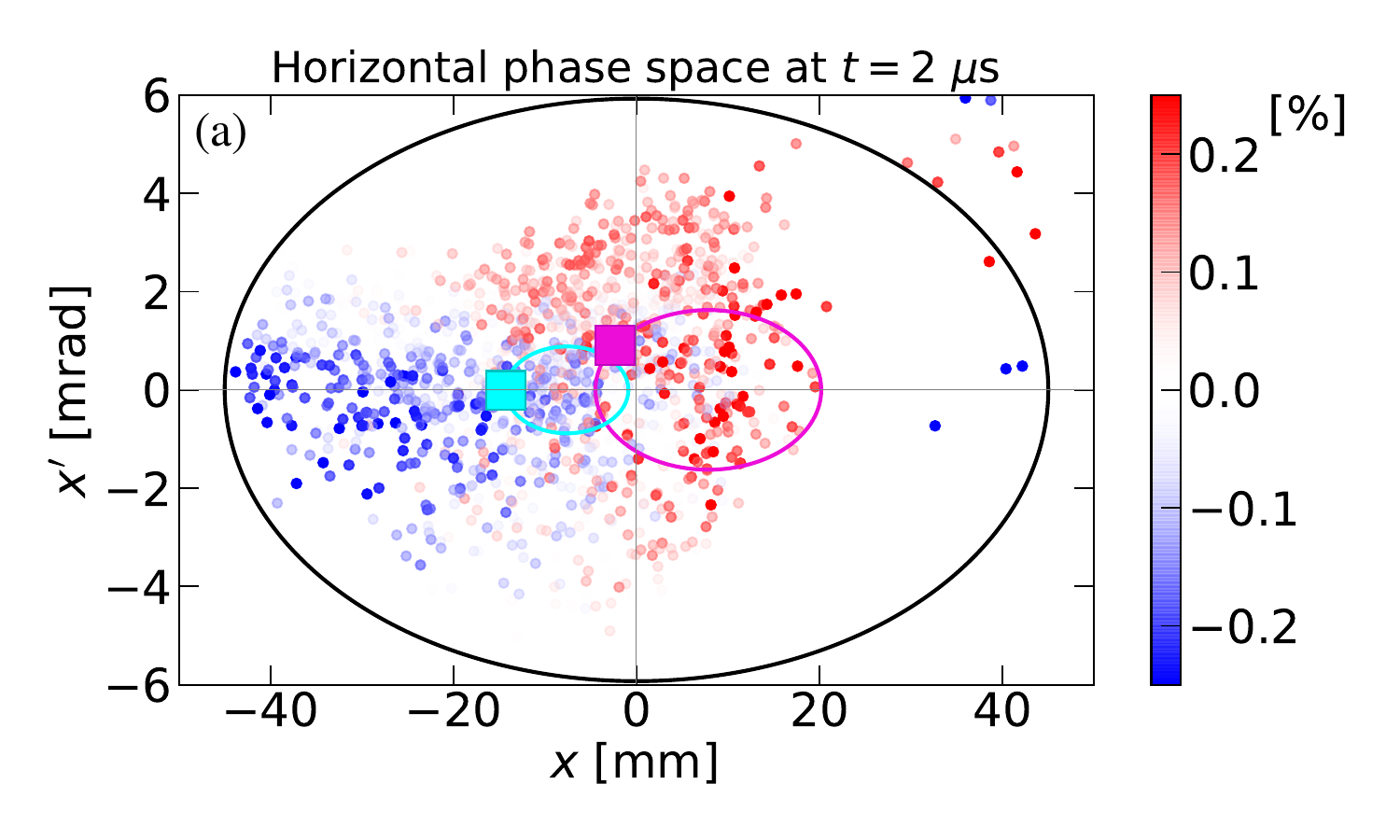}
		\label{fig:phasespace_a}
	\end{subfigure}
	~
	\begin{subfigure}{\linewidth}
		\includegraphics[width=\textwidth]{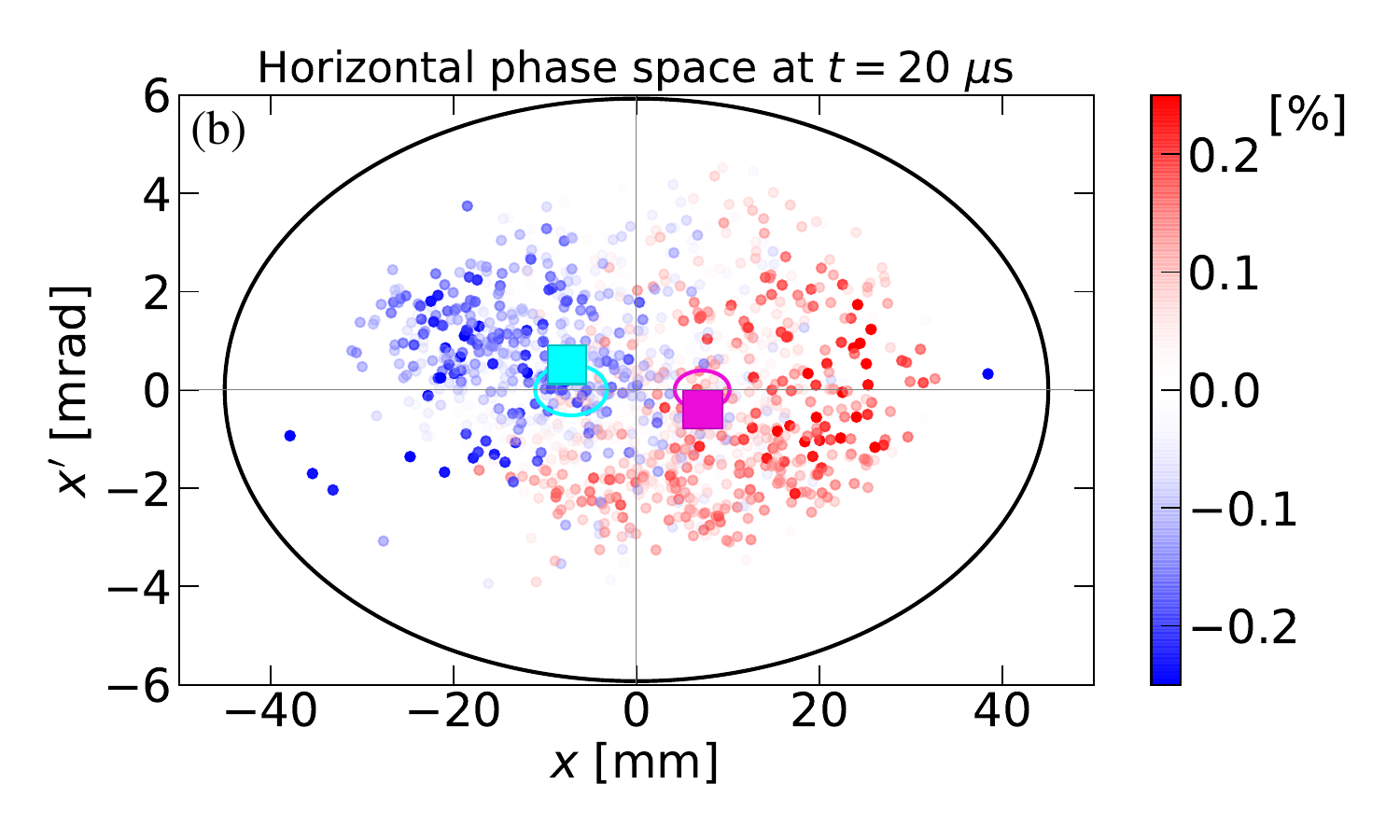}
		\label{fig:phasespace_b}
	\end{subfigure}
	~
	\begin{subfigure}{\linewidth}
		\includegraphics[width=\textwidth]{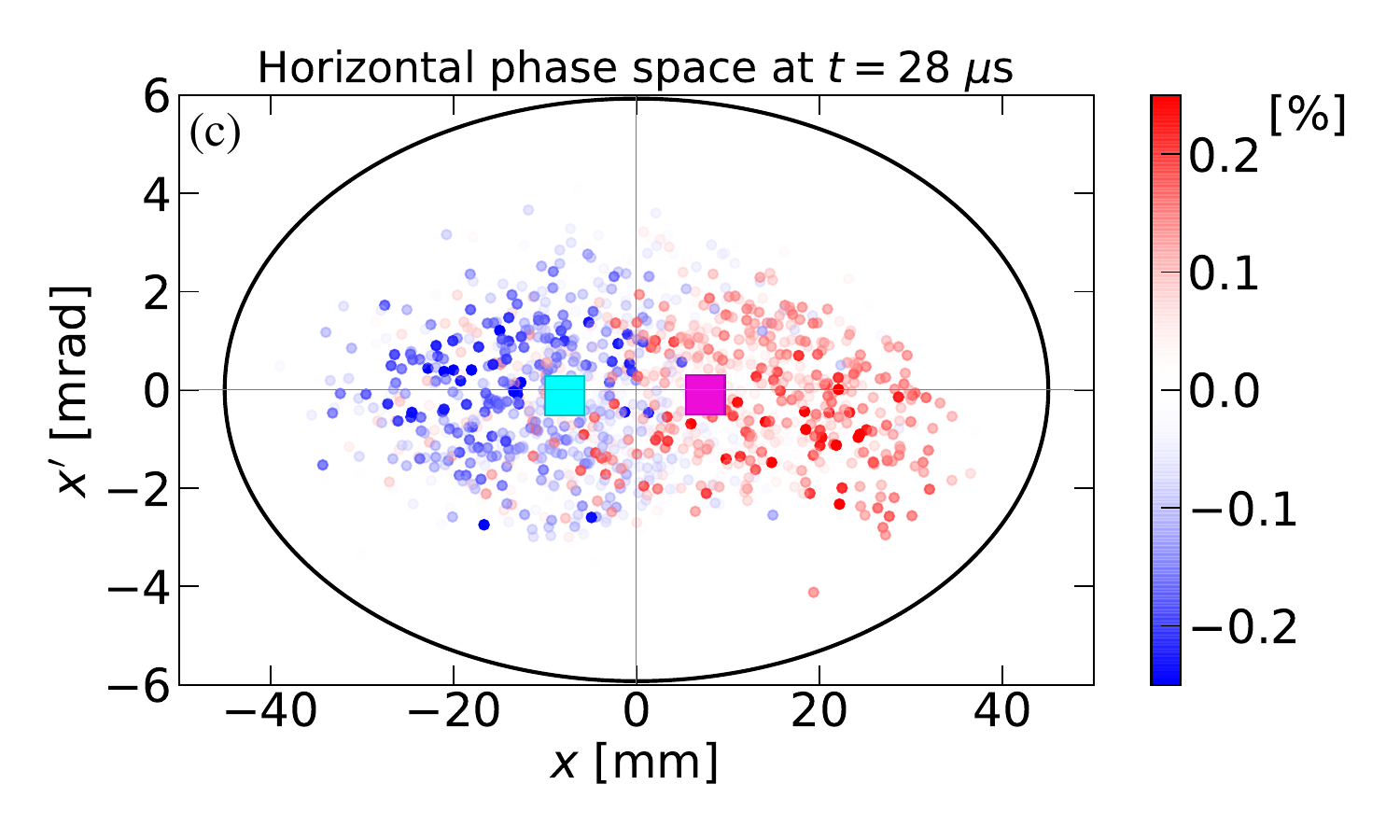}
		\label{fig:phasespace_c}
	\end{subfigure}
	\caption{Snapshots of the horizontal phase space at different times. The momentum is represented through a color-code from $\delta p/p_\text{m}=-0.3\%$ (blue) to $\delta p/p_\text{m}=+0.3\%$ (red). The purple and cyan ellipses show the phase space trajectories of the high- and low-momentum populations, respectively. (a) $t=1.9$ $\mu$s, before any RF field is applied in the simulations. (b) $t=20$ $\mu$s, when the dipole RF field is finished. (c) $t=28$ $\mu$s, when the quadrupole RF field is finished.}
	\label{fig:phasespace}
\end{figure}

The total effect of the dipole and the quadrupole RF modes in the simulation are summarized in Figure \ref{fig:phasespace}, which shows three time slices of the phase space. The center of mass (CM) for the high-momentum population is represented by a purple square, and the phase space trajectory of the CM of the high-momentum particles is  represented by a purple ellipse. The CM of the low-momentum particles is represented by a cyan box, and the phase space trajectory for this CM is shown by a cyan ellipse.  Figure \ref{fig:phasespace} (a) is before the application of the RF fields. The dipole RF mode is then applied to the beam for 20 $\mu$s, resulting in opposite phases for the high- and low-momentum populations. Thus, the centers of the two trajectories have become symmetric around the design orbit.

This opposite phase for low- and high-momentum particles provides the ideal condition for the RF quadrupole mode to shrink the phase space of the muon beam. Figure \ref{fig:phasespace} (c) shows how the RF quadrupole field shrinks the beam in phase space in the remaining 10 $\mu$s.

\section{Hardware}

The muon beam is stored for around $750 ~\mu$s with an average repetition frequency of 12 Hz. The quadrupole plates are pulsed in such a way that the at-voltage time coincides with the measurement time. The pulsed voltage is transferred to the quadrupoles through high voltage (HV) resistors with approximately $5~\mu$s RC  time constants. The E821-type scraping is also applied at this ramping period. 

The quadrupole plates and cages are the same as used in E821~\cite{ref:bnl_quad}, with the exception of the first quadrupole (Q1), where the beam enters the storage ring through the quadrupole plate. Q1 has been redesigned and  is significantly thinner than in E821.  The E989 high voltage pulsing circuit is new and will be described in Ref.~\cite{ref:e989_quads_2019}.

\begin{figure}
	\centering
	\includegraphics[width=\linewidth]{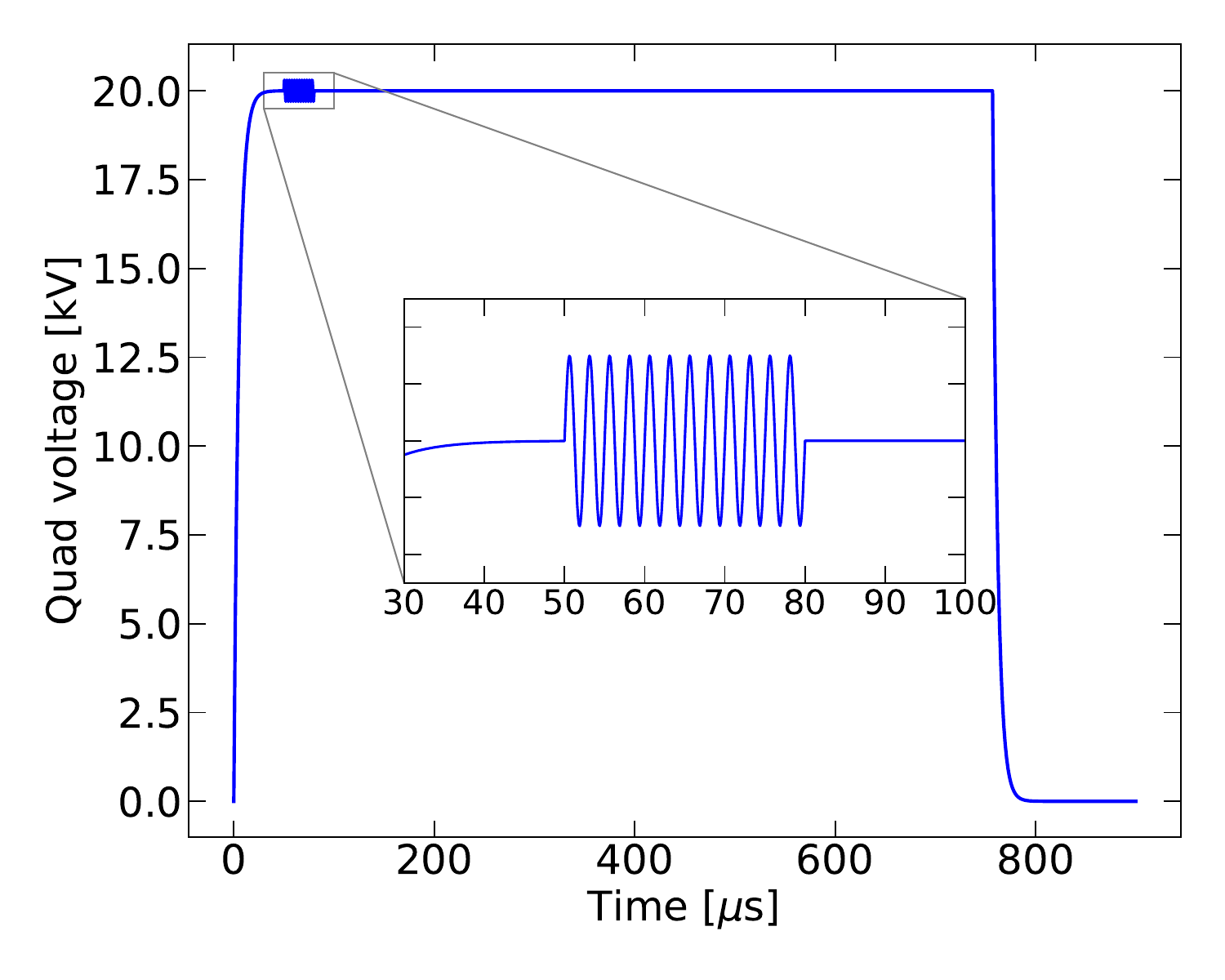}
	\caption{The RF voltage is superposed with the original HV pulse (20 kV in this example). The beam is stored for around $750~\mu$s, while the RF field is applied during a small fraction of it.  The sinusoidal signal in the figure is inserted to show the scale. During the operations, the signal is composed of three separated sinusoidals for scraping, RF dipole and RF quadrupole modes.}
	\label{fig:QuadPulseModulate}
\end{figure}

\begin{figure}
	\centering
	\includegraphics[width=0.9\linewidth]{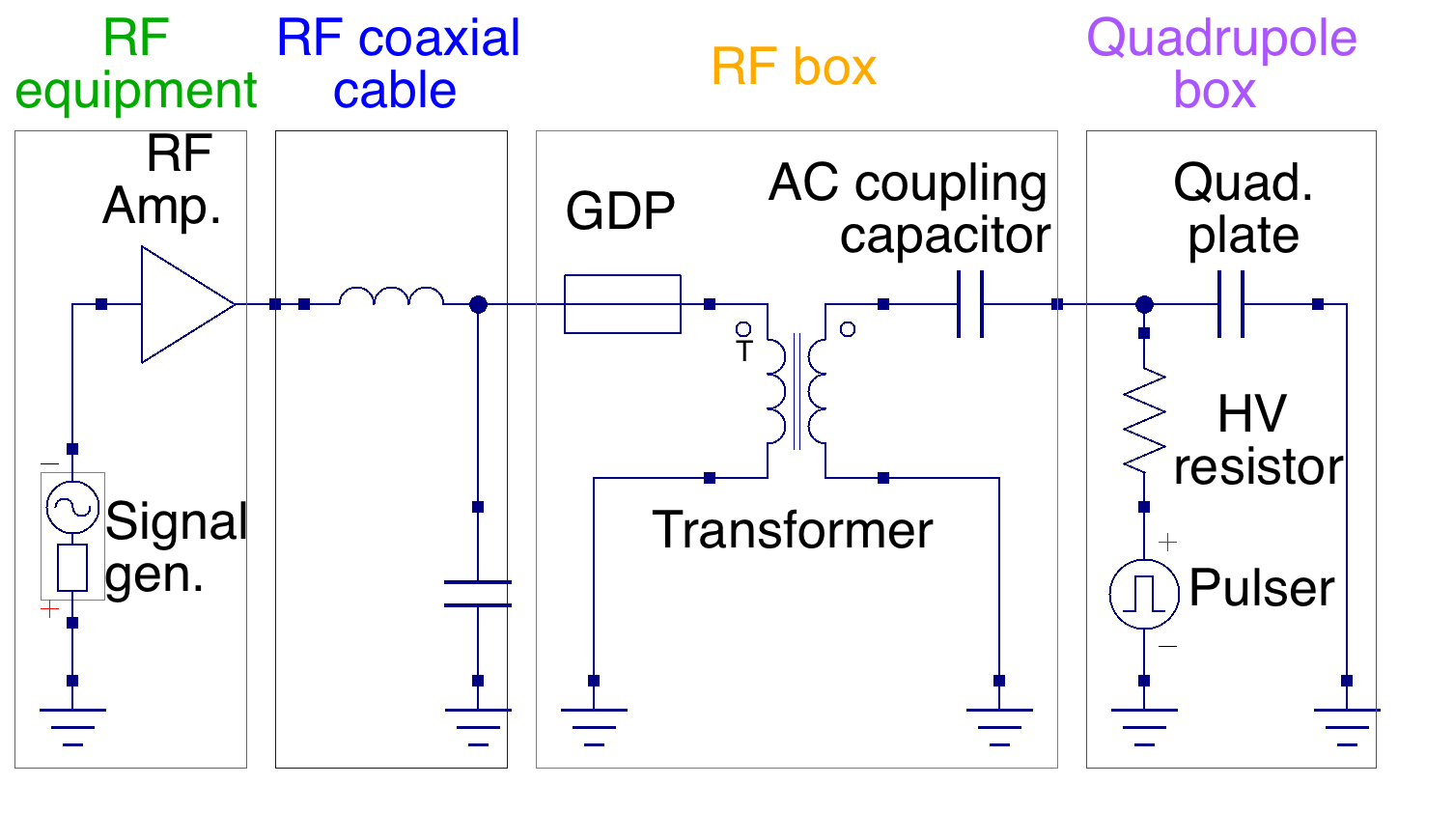}
	\caption{Circuit diagram of the RF system and original quadrupole feedthrough design. The 3D drawing of the RF box and the quadrupole box is given in Figure \ref{fig:HV_box}.}
	\label{fig:total_RF}
\end{figure}

\begin{figure}
	\centering
	\includegraphics[width=\linewidth]{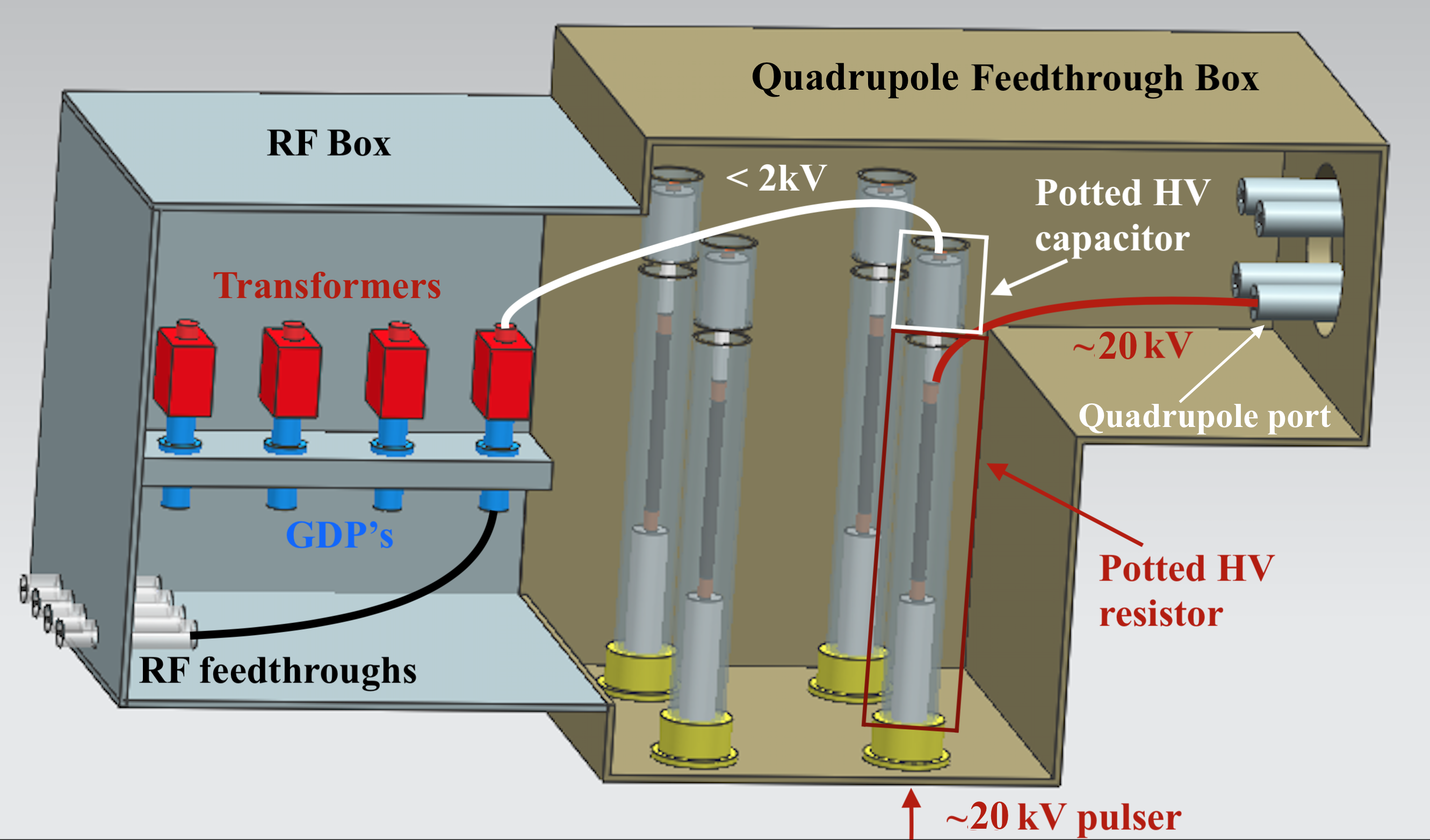}
	\caption{The interface between the quadrupoles and the RF electronics. The potted HV capacitors and the parts inside the RF box did not exist in the original design. The HV pulse ends up at the quadrupoles through the potted HV resistors, while the HV capacitor limits it to less than 2 kV at the RF feedthrough. Only one set of the cable connections is shown for clarity.}
	\label{fig:HV_box}
\end{figure}

The RF voltage is superimposed on the HV pulses (Figure \ref{fig:QuadPulseModulate}) to modulate the field at the quadrupole plates. Figure \ref{fig:total_RF} shows the circuit diagram of the system. While the RF voltage is applied through the RF electronics on the left, the main field of the quadrupole is applied through a HV resistor. The RF electronics is protected by a gas discharge protector (GDP), which is connected to a transformer for impedance matching and a potted HV capacitor that couples the RF to the quadrupole plates. Figure \ref{fig:HV_box} shows a 3D drawing of the installed system. There are four of each element in the box for each quadrupole plate. 

The RF electronics is composed of trigger gate/delay generators, fan-in/fan-out modules, 7 signal generators, 13 HV pulsed amplifiers, and HV RF cables for every quadrupole plate. The main trigger for the quadrupole pulser is used for generating a gate of 40 $\mu$s, which is copied for each signal generator and the amplifier. The signal generators are used in ``arbitrary signal mode'', and controlled by means of a GUI. For CBO reduction and scraping with dipole fields, the arbitrary function is composed of two sinusoidal signals with integer number of oscillations, which are separated by a short gap to switch the phase. Each signal has a delay before the first sine wave, which is determined by the azimuthal location of the connected quadrupole. The RF amplifiers with $\approx57$ dB gain deliver the signal through the HV RF cables to the RF box, which is depicted in Figure \ref{fig:HV_box}. The application of the RF system does not require any modification in the experimental conditions.

\begin{figure}
	\centering
	\includegraphics[width=\linewidth]{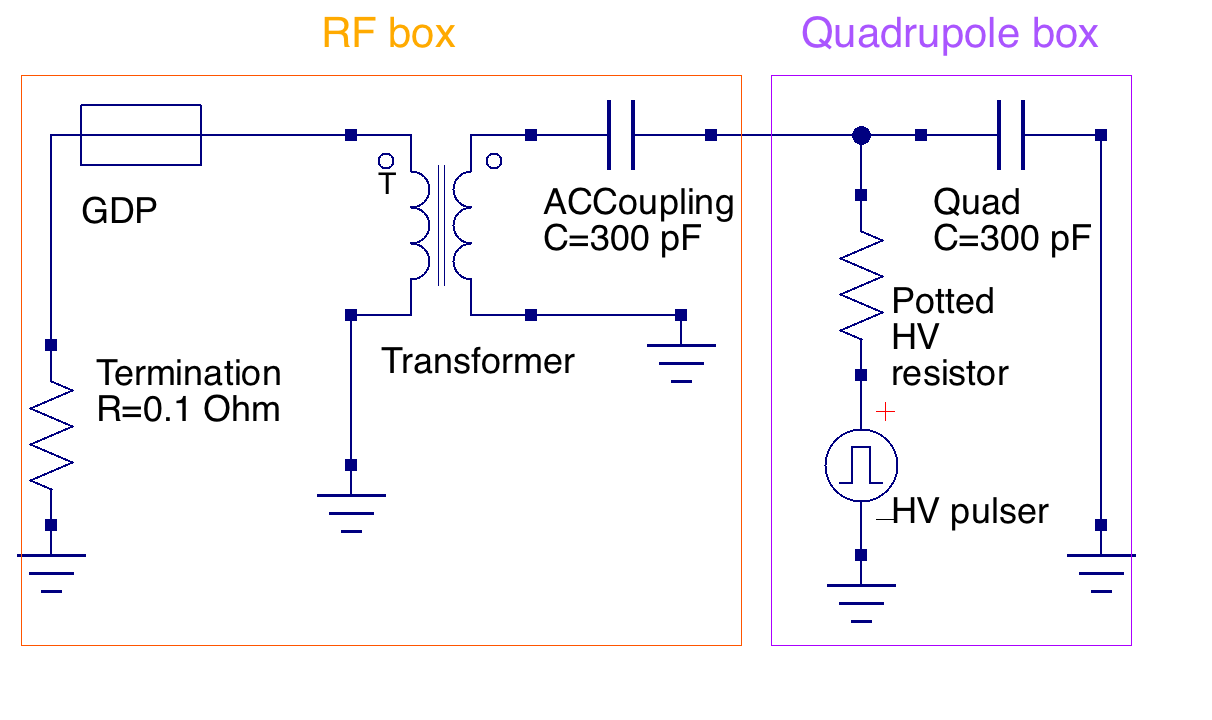}
	\caption{The HV circuit is back-compatible. The RF electronics part of Figure \ref{fig:total_RF} can be shorted to the ground  to go back to the E821-type scraping.}
	\label{fig:installed_system}
\end{figure}

As mentioned earlier, the RF reduction method has an efficient scraping mode as well, which is preferable to the conventional method since the field index does not change during scraping in this method. Nevertheless, the RF electronics can be disconnected and the RF branch of the circuit can be shorted to the ground  (shown with 0.1 Ohm resistor in Figure \ref{fig:installed_system}) for back compatibility with the original scraping scheme.

\begin{figure}[]
	\centering
	\includegraphics[width=0.7\linewidth]{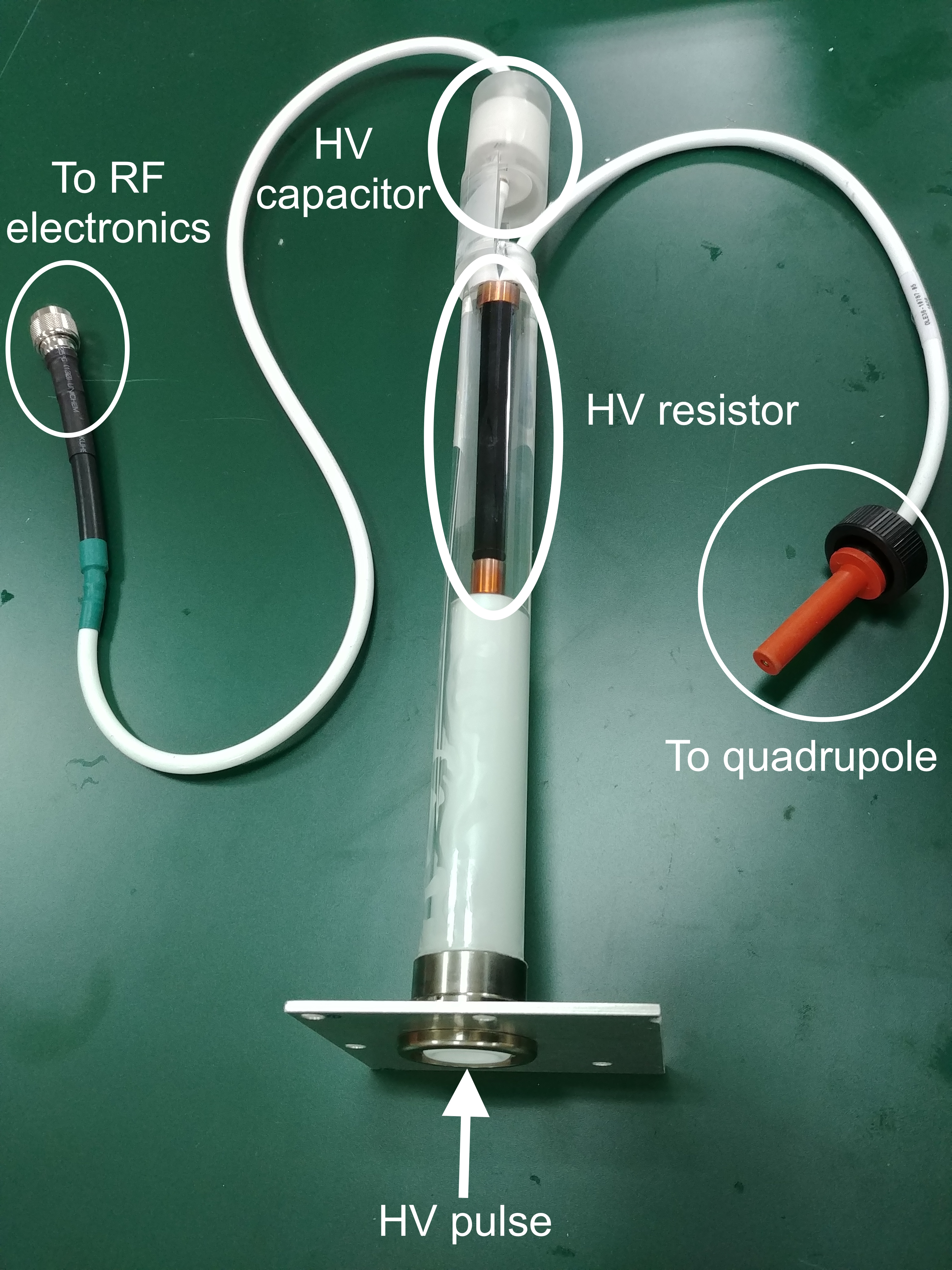}
	\caption{The resistor and the AC coupling capacitor are potted separately. The capacitor and the cable to the RF electronics did not exist in the original design. }
	\label{fig:potted}
\end{figure}

HV breakdowns are initiated by the emission of electrons from the cathode to anode at high voltages. The exchange of charged particles between the cathode and anode can be enhanced through several mechanisms~\cite{ref:breakdown1, ref:breakdown2, ref:breakdown3}. One way of avoiding electron HV breakdowns is to eliminate the contact between the air and conductors. The HV resistors and capacitors are potted in respective containers with silicone elastomer (SE) for this purpose (Figure \ref{fig:potted}). 

\section{Test results}

The hardware was tested by installing the RF system into all of the quadrupole sections in the E989 experiment. A 450 V amplitude RF voltage with an azimuthally dependent phase was applied to all of the quadrupole sections. The voltage was split between the long and the short quadrupoles by an HV divider. The motion of the beam centroid was extracted from the tracker detector \cite{ref:tracker} data. 

\begin{figure}[h!]
	\centering
	\includegraphics[width=\linewidth]{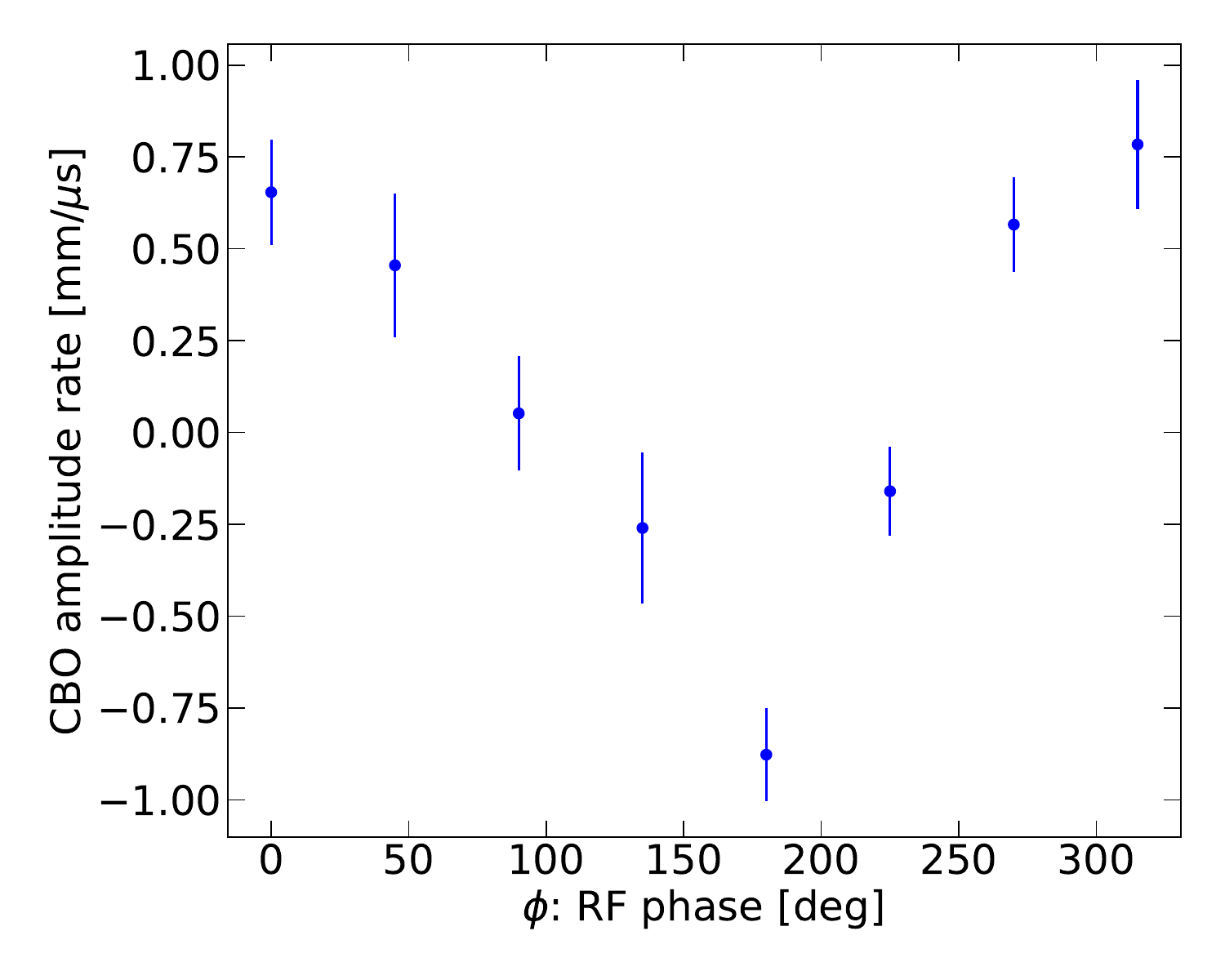}
	\caption{The CBO amplitude reduction rate with respect to the RF phase. The data were obtained with a 2-hour scanning. The maximum reduction was achieved at around $\phi=160^\circ$.}
	\label{fig:datasimulation}
\end{figure}

Figure \ref{fig:datasimulation} shows the CBO amplitude damping rate that is obtained during the phase scan. Negative and positive values indicate CBO reduction and growth, respectively. The maximum reduction happens at $\phi_0=160^\circ$. 

\begin{figure}
	\centering
	\includegraphics[width=\linewidth]{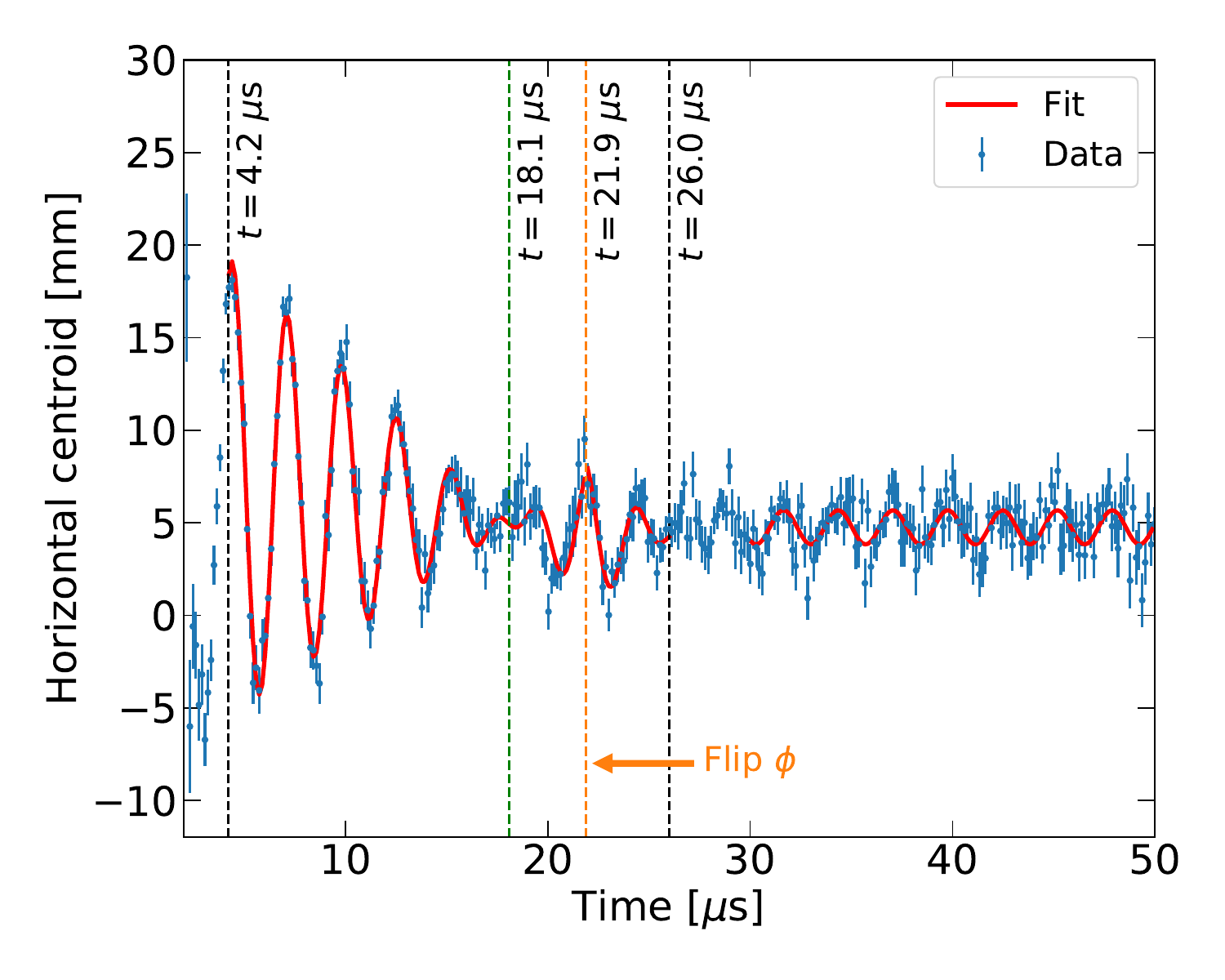}
	\caption{The tests with the dipole RF fields showed a factor of 15 reduction in $13.9~\mu$s, shown with the green dashed line. Then, the CBO amplitude increased with the opposite phase as the RF was not stopped. This feature was exploited for scraping the beam. Finally, at $t=21.9~\mu$s, the phase was flipped by $180^\circ$ and then turned off at $t=26~\mu$s. The sinusoidal fits at each epoch are shown in red.}
	\label{fig:rf_test_result_cbo_reduction}
\end{figure}

Figure \ref{fig:rf_test_result_cbo_reduction} shows the CBO measurement at the RF phase $\phi_0=160 ^ \circ$. Initially, the CBO amplitude is $\approx 15$ mm. The RF field starts at $t=4.2~\mu$s, reducing the CBO amplitude to $x\approx 1$ mm at  $t=18.1~\mu$s. Then, it grows similar to Figure \ref{fig:YuriFormulaConfirm} with the RF field. It is worth noting that the CBO phase flips by $180 ^\circ$ after the crossover. This feature was exploited for scraping the beam.

We continued the application of the RF field at the same phase for scraping the beam. Then, the RF field was flipped by $180^\circ$ at $t=21.9~\mu$s. Finally, it was turned off at $26~\mu$s. At that time, the CBO amplitude reduced back to 1 mm after scraping. The whole CBO reduction and the scraping process took around  $22~\mu$s.

The tests reveal that RF scraping produces results comparable to those of conventional scraping. Measurement of the betatron frequencies shows that the RF scraping and the CBO reduction do not affect the beam dynamics after the RF time is over.


\section{Conclusions}

This work represents the first successful development of the hardware necessary to apply RF cooling to a tertiary particle beam. Unlike primary beams in storage rings, where RF manipulation can  occur over millisecond time scales, the muon time dilated lifetime of $\gamma \tau_\mu = 64.4\ \mu$s severely limits the time available for beam manipulation, if there is to be a significant measurement time following the beam cooling period. 

The RF reduction method described here has been shown to significantly reduce the CBO while achieving loss of muons comparable to that which occurs during conventional scraping. This improvement is essential to determine the muon spin rotation frequency, and thereby the muon magnetic anomaly $a_\mu$ with an unprecedented sensitivity.

\section{Acknowledgements}
This work was supported by IBS-R017-D1 of the Republic of Korea, and also in part by Fermi Research Alliance, LLC under Contract No. DE-AC02-07CH11359 with the U.S. Department of Energy, Office of Science, Office of High Energy Physics. The U.S. university groups were supported by U.S. DOE Office of High Energy Physics research grants. We would also like to thank the many $g-2$ collaborators who have helped us with this project. We also would like to thank Sidney Orlov for improving the readability of the paper.



\begin{thebibliography}{2}
	
	\bibitem{ref:miller_2007}{J. P. Miller, E. de Rafael, and B. L. Roberts, Muon g-2: Review of Theory and Experiment, Rept. Prog. Phys. {\bf 70} 795, https://doi.org/10.1088/0034-4885/70/5/R03 (2007).}
	\bibitem{ref:bailey_1969pr} {J. Bailey \etal, Precision measurement of the anomalous magnetic moment of the muon, Phys. Lett. {\bf B28}, 287 (1968).} 
	\bibitem{ref:cern_3}{J. Bailey \etal, Precise measurement of the anomalous magnetic moment of the muon, Il Nuovo Cimento A {\bf 9},  369 (1972).} 
	\bibitem{ref:cern_4}{J. Bailey \etal, Final report on the CERN muon storage ring including the anomalous magnetic moment and the electric dipole moment of the muon, and a direct test of relativistic time dilation, Nucl. Phys. B {\bf 150}, 1 (1979).} 
	\bibitem{ref:bnl_final_report}{G.W. Bennett \etal, Final report of the E821 muon anomalous magnetic moment measurement at BNL,  Phys. Rev. D \textbf{73}, 072003 (2006).} 
	\bibitem{ref_sensitivity_1}{M. Davier \etal, Reevaluation of the hadronic  vacuum polarization contributions to the Standard Model predictions  of the muon $g-2$  and $\alpha (m_Z^2)$ using newest hadronic cross-section data, Eur. Phys. J. C \textbf{77}, 827 (2017).}  
	\bibitem{ref:sensitivity_2}{A. Keshavarzi \etal, Muon $g-2$ and $\alpha (m_Z^2)$: A new data-based analysis, Phys. Rev. D
		\textbf{97}, 114025 (2018).} 
	\bibitem{ref:fermilab}{J. Grange \etal, Muon $g-2$ Collaboration, Muon $g-2$ technical  design report, arXiv:1501.06858 (2015).} 
	\bibitem{ref:Danby_2001eh}{G. T. Danby \etal, The Brookhaven muon storage ring magnet, Nucl. Inst. and Meth. A {\bf 457} 151-174, https://doi.org/10.1016/S0168-9002(00)00704-X (2001).}
	\bibitem{ref:Yamamoto_2002bb}{A. Yamamoto \etal, The superconducting inflector for the BNL g-2 experiment, Nucl. Inst. and Meth. A {\bf 491} 23-40,  https://doi.org/10.1016/S0168-9002(02)01232-9 (2002).}
	\bibitem{ref:inflector}{N. S. Froemming \etal,  Commissioning the superconducting magnetic inflector system for the muon g-2 experiment,  IPAC'18 Proceedings,  https://doi.org/10.18429/JACoW-IPAC2018-WEPAF014 (2018).}
	
	
	\bibitem{ref:bnl_kicker}{E. Efstathiadis \etal, A fast non-ferric kicker for the muon (g-2) experiment, Nucl. Inst. and Meth. A {\bf 496} 8-25 (2003). }
	
	\bibitem{Schreckenberger:2018njd} {A.P. Schreckenberger  \etal, New Fast Kicker Results from the Muon g-2 E-989 Experiment at Fermilab, Proc. 9th International Particle Accelerator Conference,  doi  "10.18429/JACoW-IPAC2018-THPML093", http://lss.fnal.gov/archive/2018/conf/fermilab-conf-18-167-e.pdf.}
	
	\bibitem{Miller:2018jum} {J. P. Miller and B. Lee Roberts,  The Muon $(g-2)$ Spin Equations, the Magic $\gamma$, What's small and what's not,  arXiv:1805.01944v2 (2018).}
	
	\bibitem{ref:bennett_2007} {G.W. Bennett \etal, Statistical  equations and methods applied to the precision muon  
		experiment at BNL,  Nucl. Inst. and Meth.A \textbf{579}, 1096 (2007).}
	
	\bibitem{ref:cbo_analysis_1}{G.W. Bennett \etal, Measurement of the Negative Muon Anomalous Magnetic Moment to 0.7 ppm, Phys. Rev. Lett. \textbf{92}, 161802 (2004).} 
	\bibitem{Orlov:2003} Y.F. Orlov and Y.K Semertzidis, To get rid of CBO (and to get scraping without resonance crossings), Muon g-2 Note 431 (2003), unpublished.   
	
	\bibitem{ref:bnl_quad}{Y.K. Semertzidis \etal, The Brookhaven muon  ($g-2$) storage ring high voltage quadrupoles, Nucl. Inst. and  Meth. A {\bf 503}, 458 (2003).}  
	
	\bibitem{ref:fnal_quad}{J.D. Crnkovic \etal, Commissioning the Muon g-2 Experiment Electrostatic Quadrupole System, IPAC2018 Proceedings, https:// doi.org/10.18429/JACoW-IPAC2018-WEPAF015 (2018).} 
	
	\bibitem{ref:tracking_benchmarks}{E.M. Metodiev \etal, Analytical benchmarks for precision particle tracking in electric and magnetic rings, Nucl. Inst. and Meth. A {\bf 797} 311-318 (2015). } 
	
	
	\bibitem{ref:dRubin}{D. Rubin, private communications (2017).}
	\bibitem{ref:e989_quads_2019}{J.D. Crnkovic \etal, Commissioning the Electrostatic Quadrupole System for the Muon g-2 Experiment, in preparation.}         
	%
	%
	\bibitem{ref:breakdown1}{R. Latham, High voltage vacuum insulation, Academic Press (1995).} 
	
	\bibitem{ref:breakdown2}{G. Farrall, IEEE Trans. Electr. Ins., E{\bf I-20}, 815 (1985).} 
	\bibitem{ref:breakdown3}{R. Morrow and D. Weisser, Vacuum breakdown mechanisms, and X-ray pulses in accelerators, Nucl. Instr. and Meth. A, {\bf 382}, 66-72  (1996).} 
	\bibitem{ref:tracker}{J. Mott \etal, The readout system for the Fermilab Muon g-2 straw tracking detectors, ICHEP2016 Proceedings, http://doi.org/10.22323/1.282.1163 (2018).}
	
	
\end{thebibliography}
\end{document}